\documentclass[aps,prd,amssymb,eqsecnum,nofootinbib,floatfix, a4paper,twocolumn]{revtex4}

\usepackage{mathrsfs}
\usepackage{latexsym,amsmath,amsfonts,amssymb}
\usepackage{bm}

\usepackage{graphicx}

\usepackage{xcolor}  

\allowdisplaybreaks

\newcommand{\be}{\begin{equation}}
\newcommand{\ee}{\end{equation}}
\newcommand{\bea}{\begin{eqnarray}}
\newcommand{\eea}{\end{eqnarray}}

\def\e{{\rm e}}

\def\tal{{\widetilde{\alpha}}}
\def\bal{{\overline{\alpha}}}

\def\k{\kappa}
\def\lam{{\lambda}}

\def\d{\partial}
\def\l{\left(}
\def\r{\right)}

\def\t0{\tilde{0}}

\def \bY{\overline{Y}}

\def\cA{{\cal A}}
\def\cF{{\cal F}}
\def\cR{{\cal R}}

\newcommand{\bg}{\begin{gather}}
\newcommand{\eg}{\end{gather}}
\newcommand{\bseq}{\begin{subequations}}
\newcommand{\eseq}{\end{subequations}}

\begin{document}

\title{Spherically symmetric solutions in torsion bigravity}

\author{Thibault \surname{Damour}}
\author{Vasilisa \surname{Nikiforova}}
 
\affiliation{Institut des Hautes Etudes Scientifiques, 
91440 Bures-sur-Yvette, France}

\date{\today}

\begin{abstract}
We study spherically symmetric solutions in a four-parameter Einstein-Cartan-type class of theories.
These theories include torsion, as well as the metric, as dynamical fields, and contain 
only two physical excitations (around flat spacetime): 
a massless spin-2 excitation and a massive spin-2 one (of mass $ m_2 \equiv \k$). 
They offer a geometric framework (which we propose to call
``torsion bigravity") for a modification of Einstein's theory that has the same spectrum as bimetric gravity models. 
We find that the spherically symmetric solutions of torsion bigravity theories exhibit several remarkable features: (i) they have the
same number of degrees of freedom as their analogs in {\it ghost-free} bimetric gravity theories ({\it i.e.} one less
than in ghost-full bimetric gravity theories); (ii) in the limit of small mass for the spin-2 field ($ \k \to 0$), 
no inverse powers of $\k$ arise at the first two orders of perturbation theory (contrary to what happens in bimetric gravity where
$1/\k^2$ factors arise at linear order, and $1/\k^4$ ones at quadratic order). We numerically construct a high-compactness
(asymptotically flat) star model in torsion bigravity and show that its geometrical and physical properties are significantly
different from those of a  general relativistic star having the same observable Keplerian mass.
\end{abstract}

\maketitle

\section{Introduction} \label{sec1}

Einstein's theory of gravitation, i.e. General Relativity (GR), has, so far, been found to be in excellent accord with
all gravitational observations and experiments. In particular, its foundational stone, the weak equivalence principle,
has been recently confirmed at the $10^{-14}$ level \cite{Touboul:2017grn}, while gravitational-wave observations
have confirmed several basic dynamical predictions of GR \cite{LIGOScientific:2019fpa,Abbott:2018lct}.
[See, e.g., chapter 20 in Ref. \cite{Tanabashi:2018oca} for a review of the experimental tests of GR.]

However, since the discovery of GR more than a century ago, the quest for possible extensions of GR has been
going on. We shall not discuss here the various motivations underlying the study of modified theories of gravity
(see Ref. \cite{Capozziello:2011et} for a review).
Let us only mention that, from a pragmatic point of view, it is useful to have alternative theories of gravity to conceive and 
interpret tests of gravity \cite{Berti:2015itd}.

Here, we study a class of geometric theories of gravitation that generalize the Einstein-Cartan theory.
 The original idea of Cartan \cite{Cartan:1923zea,Cartan:1924yea,Cartan1925} was to extend GR by considering 
 the metric and the (affine) connection as {\it a priori}  independent fields (first-order formalism), and by allowing the
 connection\footnote{Cartan worked within a vielbein formalism, in which the affine connection
 is naturally restricted to be metric-preserving; see below.} to have nonzero torsion.  Cartan added the idea that  torsion might be sourced by some sort of intrinsic spin
 density along the matter worldlines\footnote{``En admettant la possibilit\'e d' \'el\'ements de mati\`ere dou\'es de
 moments cin\'etiques non infiniment petits par rapport \`a leur quantit\'e de mouvement."; bottom of p. 328 in \cite{Cartan:1923zea}}. Later, Weyl pointed out that it is natural, in such a first-order formalism, to consider that fermions (Dirac spinors)
 directly couple to the connection, so that the torsion  ${T^i}_{[jk]}= - {T^i}_{[kj]}$ is sourced by the microscopic (quantum) spin density of fermions
 $\sim \frac12 \bar \psi \gamma^i \gamma_{[j} \gamma_{k]} \psi$. [As explained in detail below, the latin indices $i, j, k, \cdots$
 denote frame indices.]
 He also showed that if one follows Einstein and Cartan
 in using as gravitational action the first-order form of the scalar curvature, the torsion is algebraically determined by its source
 and that the first-order action is equivalent to a second-order (purely metric) action containing additional ``contact terms"
 quadratic in the torsion source, and therefore quartic in fermions. 
 The ideas of Cartan and Weyl were further developed by Sciama \cite{Sciama1962}, Kibble \cite{Kibble:1961ba} and 
 many others (see \cite{Hehl:1976kj} for a review of later work on this approach based
 on gauging the Poincar\'e group).
 
 A new twist in the story started after the discovery of supergravity \cite{Freedman:1976xh}, and especially of its first-order formulation \cite{Deser:1976eh}. Indeed,  the first-order formulation of supergravity is similar to the Einstein-Cartan-Weyl
 approach, with a gravitational term linear in the scalar curvature,
 and a nonzero torsion ${T^i}_{[j k]}$ algebraically determined in terms of its gravitino source
 $\sim  \bar \psi_j \gamma^i  \psi_k$, leading, after replacement in the action, to contact terms quartic in the gravitino.
 However, quantum loops generate an effective action containing terms at least quadratic in the curvature. When considered
 in a purely metric, second-order formulation, terms quadratic in the curvature lead to higher-order field equations,
 which raise difficulties \cite{Stelle:1976gc}, in the form of ``ghosts" (negative-energy modes), 
 even at the classical level \cite{Stelle:1977ry}. This raised the issue
 of finding ghost-free theories of gravity with an action containing terms quadratic in curvature,
 but treated in a first-order formulation. Indeed, such a formulation
  leads to second-order-only field equations for the metric and the connection
 \cite{Neville:1978bk,Neville:1979rb}, so that the torsion now propagates away from its source. 
 The most general solution to finding such ghost-free and tachyon-free (around Minkowski spacetime)
 theories with propagating torsion was obtained in parallel work by Sezgin and van Nieuwenhuizen \cite{Sezgin:1979zf,Sezgin:1981xs}, and by Hayashi and Shirafuji \cite{Hayashi:1979wj,Hayashi:1980av,Hayashi:1980ir,Hayashi:1980qp}.
 It was found that there are twelve six-parameter families of ghost-free and tachyon-free theories with propagating torsion
 \cite{Hayashi:1980qp,Sezgin:1981xs}. These theories always contain an Einsteinlike massless spin-2 field,
 together with some (generically) massive excitations coming from the torsion sector. The possible spin-parity labels
 of the excitations propagated by the torsion sector are: $2^+$, $2^-$, $1^+$, $1^-$, $0^+$, $0^-$. Only certain combinations
 of these spin-parities can be present in the various six-parameter families of ghost-free and tachyon-free theories with propagating torsion (see Table I in \cite{Hayashi:1980qp} or Table I in \cite{Sezgin:1981xs}).

 One of these classes of theories (with torsion propagating both  massive $2^+$ and massive $0^-$ excitations)
 has recently been studied with the hope
 that the massive spin-2 field it contains will define a new, more geometric, solution 
 to having a healthy and cosmologically relevant infrared modification of gravity \cite{Nair:2008yh,Nikiforova:2009qr}.
 We recall that the physics of an ordinary, massive\footnote{especially with a very small mass, say of cosmological scale.} 
 Fierz-Pauli-type \cite{Fierz:1939zz,Pauli:1939xp,Fierz:1939ix} spin-2 field raises many subtle issues going by the names of:
 vanDam-Veltman-Zakharov discontinuity \cite{vanDam:1970vg,Zakharov:1970cc}, 
 Vainshtein (conjectured) mechanism \cite{Vainshtein:1972sx}, and Boulware-Deser ghost \cite{Boulware:1973my}.
 A breakthrough in the problem of defining a class of consistent, ghost-free nonlinear theories of a massive spin-2 field
 was achieved in Ref. \cite{deRham:2010kj}. This then allowed the construction of a class of consistent, ghost-free nonlinear 
  theories of {\it bimetric gravity} \cite{Hassan:2011zd}.
  
 The aim of the present paper is to study the four-parameter subclass of the propagating-torsion models of Refs. 
 \cite{Sezgin:1979zf,Sezgin:1981xs,Hayashi:1979wj,Hayashi:1980av,Hayashi:1980ir,Hayashi:1980qp,Nair:2008yh,Nikiforova:2009qr} that it similar to the bimetric gravity models of \cite{Hassan:2011zd} in the sense
 that it contains only two types of excitations: an Einsteinlike massless spin-2 excitation, and a positive-parity massive spin-2 one.
 To emphasize this similarity we shall often refer to the models we study as defining a theory of {\it torsion bigravity}.
 We think that the geometric origin of the massive spin-2 additional field (contained among the torsion components, rather than
 through a second metric) makes such a torsion bigravity model an  attractive alternative to the usually
 considered bimetric gravity models. In particular, the fact that massive gravity is described in these
 models by a different Young tableau than  the more familiar (symmetric tensor) models completely changes the
 various issues linked to nonlinear effects, and renders the study of their
 physical properties {\it a priori} interesting. Some of the results of previous work on such models 
 \cite{Nair:2008yh,Nikiforova:2009qr,Nikiforova:2016ngy,Nikiforova:2017saf,Nikiforova:2017xww,Nikiforova:2018pdk}
  has shown them to be remarkably healthy and robust around various backgrounds (though Ref. \cite{Nikiforova:2018pdk}
  found the presence of gradient instabilities around the self-accelerating torsionfull cosmological solution found in 
  \cite{Nikiforova:2016ngy}; but these instabilities might be due to the endemic stability problems of 
  self-accelerating cosmological universes rather than to the theory itself). Anyway, let us emphasize here that
  the existence of the self-accelerating solution of Ref. \cite{Nikiforova:2016ngy} necessarily relied on the
  presence in the spectrum of {\it both} $2^+$ and $0^-$ excitations. In the present work we focus on the
  minimal model containing {\it only} the $2^+$ excitation (besides the Einstein massless graviton). This
  minimal torsion bigravity model has not yet received any specific attention in the literature beyond its linearized
  approximation (which follows from the general linearized-limit results of Refs. \cite{Hayashi:1980ir,Zhang:1982jn,Nikiforova:2009qr}).

 Let us note in passing, for the cognoscenti, that we are talking here about 
 positive-parity spin-2
 excitations contained in the torsion field ${T^i}_{[j k]}$, and not of the ``dual gravity", negative-parity spin-2 excitation 
contained in the irreducible SO$(3,1)$ Young tableau $T_{[ab]c}$ (satisfying $T_{[ab]c}+ T_{[bc]a}+T_{[ca]b}=0$)
introduced by Curtright \cite{Curtright:1980yk,Curtright:1980yj}.  Among the propagating torsion models of  
Refs. \cite{Hayashi:1980qp,Sezgin:1981xs} some give rise to   massive $2^-$ excitations and some to   massive $2^+$ ones,
but the two parities cannot be simultaneously present in ghost-free models.

As we started this Introduction by recalling that the source of torsion is the microscopic (quantum) spin of elementary fermions,
the reader might worry that this would prevent the existence of phenomenologically relevant, macroscopic torsion fields in ordinary,
non spin-polarized systems, such as stars, planets, or even neutron stars\footnote{We leave to future work a study
of the amount of spin-polarization in a strongly magnetized neutron star.}. However, as was already noticed in Refs. 
\cite{Hayashi:1980ir} and \cite{Nikiforova:2009qr}, and as will be clear in the present work, the mere presence of a
usual, Einsteinlike energy-momentum tensor $T^{\mu \nu}$ suffices to generate macroscopic torsion fields. 
In the following, we shall then, for simplicity, set the torsion source to zero and only consider the effect of the 
energy-momentum source $T^{\mu \nu}$.

\section{Formalism and action of torsion bigravity} \label{sec2}

Here, we essentially follow the notation of Refs. \cite{Hayashi:1979wj,Hayashi:1980av,Hayashi:1980ir,Hayashi:1980qp} 
(which we also used in our previous paper  \cite{Nikiforova:2018pdk}). 
Latin indices $i, j, k, \ldots= 0,1,2,3 $ (moved
by the Minkowski metric $\eta_{ij}, \eta^{ij}$) are used to denote Lorentz-frame indices
referring to a vierbein ${e_i}^\mu$ (with inverse $ {e^i}_\mu$) , while Greek indices $\mu, \nu, \ldots=0,1,2,3$ 
(moved by the metric $g_{\mu \nu} \equiv \eta_{ij} {e^i}_\mu {e^j}_\nu$)
are used to denote spacetime indices
linked to a coordinate system $x^\mu$. When there is a risk of confusion we add a hat, e.g. ${e^{\hat i}}_\mu$, 
on the frame indices. The signature is mostly plus. 

The (first-order) action is expressed in terms of two basic independent fields: 
(i) the (inverse) vierbein $e^i_{\mu}$; and (ii) a general SO$(3,1)$ connection 
${A^i}_{ j \mu}$, which is {\it metric-preserving} (i.e. $A_{i j \mu}= - A_{j i \mu}$, where $A_{i j \mu}\equiv \eta_{is} {A^s}_{ j \mu}$).
The most general ghost-free and tachyon-free (around Minkowski spacetime) action containing only a massless spin-2 excitation and
a (positive-parity) massive spin-2 one has four parameters\footnote{See Appendix \ref{appB} for a discussion,
and the link with our previous notation.} and can be written as:
\be
S_{\rm total} = S_{\rm TBG}[{e^i}_\mu, A_{i j \mu}]+ S_{\rm matter} .
\ee
The torsion bigravity part, $ S_{\rm TBG}$, of the action reads
\be
S_{\rm TBG}[{e^i}_\mu, A_{i j \mu}] =\int d^4 x \, \sqrt{g} \,  L_{\rm TBG}[e, \d e, \d^2 e, A, \d A]\,,
\ee
where $\ \sqrt{g} \equiv  \sqrt{ - \det g_{\mu \nu}} \equiv \det {e^i}_\mu$, and
\bea  \label{lagrangian}
L_{\rm TBG} &=& c_R \, R[e, \d e, \d^2 e] +c_F \, F[e, A, \d A] \\ \nonumber
&+& c_{F^2}\left( F_{(ij)} F^{(ij)} - \frac13 F^2 \right)
 + c_{34} F_{[ij]} F^{[ij]}  \; . 
 \eea
 Here, we use the letter $R$ to denote the various curvatures defined by the Riemannian structure
 (curvature tensor ${R^i}_{jkl} \equiv  {R^i}_{ j \mu \nu} {e_k}^\mu {e_l}^\nu$, Ricci tensor 
 $R_{ij}={R^k}_{ikj}$ and curvature scalar $R=\eta^{ij}R_{ij}$), and the letter $F$
  to denote the corresponding Yang-Mills-type curvatures defined by the SO$(3,1)$ connection 
${A^i}_{ j \mu}$ ((curvature tensor ${F^i}_{jkl} \equiv  {F^i}_{ j \mu \nu}(A) {e_k}^\mu {e_l}^\nu$, Ricci tensor 
 $F_{ij}(A)= {F^k}_{ikj}$ and curvature scalar $F(A)=\eta^{ij}F_{ij}$). Note that, because of the projections
 on the frame, the frame components of the $F$-type curvature depend algebraically on the vierbein ${e^i}_\mu$, besides
 depending on ${A^i}_{ j \mu}$ and its first derivatives.
 See Appendix \ref{appA} for more details on the definition
of these objects, and for the relation with the notation used in our previous paper \cite{Nikiforova:2018pdk}.
[An explicit form of the general field equations can also be found in the latter reference.]

The torsion bigravity lagrangian \eqref{lagrangian} {\it a priori} depends on four parameters: $c_R$,
$c_F$, $c_{F^2}$, and $c_{34}$. Actually, the last one, $c_{34}$, will not enter in the discussion
of spherically symmetric solutions.
This leaves us with three relevant parameters. The analysis of 
Refs. \cite{Hayashi:1980qp,Sezgin:1981xs} has shown that the absence (around a Minkowski background)
of pathologies (ghosts or tachyons) require the three parameters $c_R$, $c_F$, $c_{F^2}$ to be {\it positive}.
Actually, they are related to the gravitationlike coupling constants $G_0$ (linked to massless spin-2 exchange)
and $G_m$ (linked to massive spin-2 exchange), and to the mass\footnote{Here, the ``mass", $\k$, of the massive
spin-2 field refers to the inverse of its (reduced) Compton wavelength, i.e. the parameter entering the exponential decay
$\propto \e^{- \k r}$ of a static torsion field.} $\k \equiv m_2$ of the massive spin-2 excitation, by the relations
\bea \label{cRcFcF2}
c_R + c_F &\equiv & \lam = \frac{1}{16 \pi G_0} \nonumber \\
\frac{c_F}{c_R} &\equiv & \eta = \frac{3}{4} \frac{G_m}{G_0} \nonumber \\
c_{F^2} &=& \frac{\eta \, \lam}{\k^2} = \frac{ c_F (1 + \frac{c_F}{c_R})}{\k^2}
\eea
Here, we have introduced (following \cite{Sezgin:1979zf}) the notation $\lam$ for the sum $c_R + c_F$ of the 
two curvature coefficients. It is indeed this sum which measures (at least in the weak field limit) the usual
Einsteinian gravitational coupling constant $1/(16 \pi G_0)$. We have also introduced the notation
$\eta$ for the dimensionless ratio $c_F/c_R$, which measures (within a factor $\frac43$ linked to the 
difference between the massless, $S_0^{\mu \nu}=T^{\mu \nu} - \frac12 T \eta^{\mu \nu}$,
and massive, $S_m^{\mu \nu}=T^{\mu \nu} - \frac13 T \eta^{\mu \nu}$, spin-2 matter couplings\footnote{In the Newtonian
limit, we have, indeed, $S_0^{00} = \frac12 T^{00}$ while $S_m^{00} = \frac23 T^{00}= \frac43 S_0^{00}$.}) 
the ratio of couplings to matter. It is tempting to conjecture that, for general solutions, the ultraminimal
class of theories defined by the three parameters $G_0$, $G_m$ and $\k = m_2$,
taking  $c_{34}=0$,  will have the best possible nonlinear behavior.

The difference between the affine connection ${A^i}_{ j \mu}$ and the torsionless Levi-Civita
connection $ {\omega^i}_{j\mu}(e)$ defined by the vierbein $e^i_{\mu}$ is called the contorsion tensor
\be \label{KvsAe}
{K^i}_{ j \mu} \equiv {A^i}_{ j \mu} - {\omega^i}_{j\mu}(e).
\ee
The frame components ${K^i}_{ j k} \equiv {e_k}^\mu {K^i}_{ j \mu}$
of the contorsion tensor are related to the frame components ${T^i}_{[jk]}= - {T^i}_{[kj]}$ of the torsion tensor by the relations
\bea
K_{ijk}&=& \frac12 (  T_{i [jk]}+ T_{j [ki]} - T_{k[ij]}) \,, \nonumber\\
T_{i [jk]} &=&  K_{ijk}- K_{ikj}.
\eea
[Note that  ${T}_{i [jk]}= - {T}_{i [kj]}$ while ${K}_{ijk}= - {K}_{jik}$.]
The field equations are linear in the second-order derivatives of $e^i_{\mu}$ and 
${A^i}_{ j \mu}$ when using these quantities as basic fields in the action. One should avoid to use the
vierbein and the torsion as basic fields because this introduces, in view of the link \eqref{KvsAe} which involves
first derivatives of the vierbein, third derivatives of the vierbein in the field equations. One should rather consider
the torsion as a field that is {\it a posteriori} derived from the basic fields.

Let us emphasize that the first-order formalism used in the  Einstein-Cartan(-Weyl-Sciama-Kibble) theory considered here
(which is often called ``Poincar\'e gauge theory") is fundamentally different from the often considered Palatini-type 
(``metric-affine") first-order formalism. In both formalisms one independently varies the metric and the connection,
and one {\it a priori} allows for the presence of torsion, i.e. for a nonsymmetric part of the connection:
\be
{T^{\lam}}_{\mu \nu} \equiv {\Gamma^{\lam}}_{\mu \nu} - {\Gamma^{\lam}}_{\nu \mu} \equiv 2\, {\Gamma^{\lam}}_{[\mu \nu]} 
\ee
However, in the Palatini approach (which is usually performed in a coordinate frame) one independently varies 
all the components of a (symmetric metric) $g_{\mu \nu}$ and of a (non-symmetric) connection ${\Gamma^{\lam}}_{\mu \nu}$.
This yields $10$  equations obtained by varying $g_{\mu \nu}$ together with $4^3=64$ equations obtained by varying the connection ${\Gamma^{\lam}}_{\mu \nu}$. By contrast, in the Cartan-type approach used here, one gets $16$ 
 equations by varying $e^i_{\mu}$ and only $24$ equations by varying $A_{i j \mu}= - A_{j i \mu}$.
Because of the (chosen) Local-Lorentz invariance of the action the $16$ vierbein equations are submitted to $6$ 
Noether identities (linked to infinitesimal local Lorentz rotations $\omega_{[ij]}$; see, e.g., 
\cite{Hehl:1976kj,Hayashi:1979wj,Nikiforova:2017saf}) and are therefore essentially equivalent to $10$ field equations 
obtained by varying $g_{\mu \nu}$. By contrast, the $64$ connection equations of the Palatini approach
are stronger than the $24$ equations obtained by varying $A_{[i j] \mu}$. For instance, if the connection
does not directly couple to matter, it has been shown \cite{Afonso:2017bxr,BeltranJimenez:2017doy} that a general Palatini action of the $\sqrt{g} f(\cR_{(\mu \nu)})$
type (where $\cR_{(\mu \nu)}$ denotes the {\it symmetric} part of the Ricci tensor defined by the
nonsymmetric connection ${\Gamma^{\lam}}_{\mu \nu}$) yields  algebraic equations for the connection 
that determine it (modulo an additional ``projective" term $\delta^\lam_{\mu} A_{\nu}$) to be
the torsionless Levi-Civita connection of the auxiliary gothic metric 
$\sqrt{q} q^{\mu \nu} \equiv \delta \left[\sqrt{g} f(\cR_{(\mu \nu)})\right]/ \delta \cR_{(\mu \nu)}$. As the projective term
drops out of the action (because it does not contribute to $\cR_{(\mu \nu)}$ and is assumed not to couple 
directly to matter) one ends up with a theory of gravity where the metric $q_{\mu \nu}$ is an Einstein-frame metric
having the usual Einstein-Hilbert dynamics, but where the matter is coupled to the different metric $g_{\mu \nu}$,
with some nonlinear relation between these two metrics and the matter stress-energy tensor $T_{\mu \nu}$.
In  these theories, there are no dynamical effects linked to a propagating torsion.
On the other hand, in the generalized Cartan-type theories considered here, the torsion field is a dynamical field,
which is generated by the matter stress-energy tensor $T_{\mu \nu}$
 even in absence of direct coupling of the connection to matter, which propagates away from the material sources, and 
 which has physical effects via its coupling to the physical metric $g_{\mu \nu}$.
 
 From the technical point of view, the crucial difference between the Cartan-type and Palatini-type approaches is that
 the  SO$(3,1)$ connection $A_{[i j] \mu}$ is algebraically constrained to be metric-preserving. This means that,
 in order to derive the Cartan-type field equations within a coordinate-based Palatini approach one needs to
 add to the action  density a Lagrange multiplier term , say
 \be
\int d^4x \, \Lambda^{\lam (\mu \nu)}  Q_{\lam ,(\mu \nu)}
 \equiv \int d^4x \, \Lambda^{\lam \mu \nu} \nabla_{\lam}^{\Gamma} g_{\mu \nu}
  \ee
  where $ Q_{\lam ,(\mu \nu)} \equiv \nabla_{\lam}^{\Gamma} g_{\mu \nu}$ denotes the covariant
  derivative of the metric with respect to the general ( a priori nonsymmetric)  affine connection ${\Gamma^{\lam}}_{\mu \nu}$.
  Note that the presence of this term in the action then contributes to the $64$ equations obtained by varying the connection
  by additional terms involving the $40$ unknown Lagrange multipliers  $\Lambda^{\lam (\mu \nu)}$.

\section{Static spherically symmetric metrics and connections}

In the present paper, we investigate static spherically symmetric solutions of torsion bigravity. We assume from the beginning
that the solutions are: (i) time-reversal invariant; (ii) SO(3) invariant; and (iii) parity invariant. Under these assumptions, we can  
use a Schwarzschildlike radial coordinate, and denote
\begin{eqnarray}
e^{2\Phi}&\equiv&-g_{00} \\
e^{2\Lambda}&\equiv&g_{rr} \;,
\end{eqnarray}
so that the metric reads:
\be \label{ds2}
ds^2=-e^{2\Phi}dt^2 + e^{2\Lambda}dr^2 + r^2\left( d\theta^2+\sin^2\theta\, d\phi^2 \right) \; \;.
\ee
We then correspondingly define the co-frame $e^{\hat{i}}= {e^{\hat{i}}}_\mu dx^\mu$ as
\be \label{frame}
e^{\hat{0}}=e^{\Phi}dt \;, \quad e^{\hat{1}}=e^{\Lambda}dr \;, \quad e^{\hat{2}}=r d\theta \;, \quad e^{\hat{3}}= r \sin\theta d\phi \;.
\ee
The structure of a general (possibly torsionfull) connection under the just stated assumptions (i)--(iii) has been determined
by Rauch and Nieh \cite{Rauch:1981tva}. This structure is clear when using Cartesianlike coordinates $x^0, x^a$, 
(with $a=1,2,3$) and a corresponding Cartesianlike co-frame $e^{\hat{0}}, e^{\hat{a}}$. 
Time-reversal invariance implies that the only nonvanishing components of a general connection must form a vector
$A_{\hat{a} \hat{0} \hat{0}}= - A_{\hat{0} \hat{a} \hat{0}}$ and a three-index tensor 
$A_{\hat{a} \hat{b} \hat{c}}= - A_{\hat{b} \hat{a} \hat{c}}$. Then spherical symmetry implies
that the vector  $A_{ \hat{a} \hat{0} \hat{0}}$ must be in the radial direction $n^{a}$, say
\be
A_{\hat{a} \hat{0} \hat{0}} = {\overline V}(r) n_{a}\,,
\ee
with some radial function ${\overline V}(r)$, while spherical symmetry, and parity invariance 
(which forbids the presence of the Levi-Civita tensor $\epsilon_{\hat{a} \hat{b} \hat{c}}$) imply
that the three-index tensor $A_{[\hat{a} \hat{b}] \hat{c}}$ must be of the form
\be
A_{[\hat{a} \hat{b}] \hat{c}} = {\overline W}(r) \left( n_{a} \delta_{bc}- n_{b} \delta_{ac} \right)\,,
\ee
with a second radial function ${\overline W}(r)$. Therefore the most general affine connection (under the assumptions (i)--(iii))
involves two {\it a priori} unknown radial functions. When re-expressing these results in terms of the polar-type frame \eqref{frame}, one finds that the two unknown radial functions parametrizing a general
affine connection can be chosen as being
\begin{eqnarray}
V(r)&=&{A^{\hat{1}}}_{\hat{0}\hat{0}}=+{A^{\hat{0}}}_{\hat{1}\hat{0}} \,,  \\
W(r)&=&{A^{\hat{1}}}_{\hat{2}\hat{2}}={A^{\hat{1}}}_{ \hat{3}\hat{3}}=-{A^{\hat{2}}}_{\hat{1}\hat{2}} = -{A^{\hat{3}}}_{\hat{1}\hat{3}}\;.
\end{eqnarray}
Note that $V$ and $W$ are components along our basic orthonormal frame \eqref{frame}.

Then the nonvanishing components of the connection one-form are found to be
\begin{eqnarray} \label{ACompts}
{A^{\hat{1}}}_{\hat{0}}&=& + {A^{\hat{0}}}_{\hat{1}}= V(r) \, e^{\hat 0}  \nonumber \\
{A^{\hat{1}}}_{\hat{2}}&=&-{A^{\hat{2}}}_{\hat{1}}=W(r) \, e^{\hat 2 } \nonumber \\
{A^{\hat{1}}}_{\hat{3}}&=& -{A^{\hat{3}}}_{\hat{1}} = W(r) \, e^{\hat 3 } \nonumber \\
{A^{\hat{2}}}_{\hat{3}}&=& -{A^{\hat{3}}}_{\hat{2}} = - r^{-1}\cot\theta \, e^{\hat 3 } 
\end{eqnarray}
Note that the last component (in the $\theta, \varphi$ 2-plane) is independent of the unknown functions $V, W$,
but only depends on the use of a polar-type frame, with a Schwarzschildlike radial coordinate.

 The nonzero components of the torsionless Levi-Civita connection one-form,
$ {\omega^i}_{j\mu}(e)$, defined by the metric \eqref{ds2}, are found to be (using Eq. \eqref{omega=C})
\begin{eqnarray} \label{OmCompts}
{\omega^{\hat 1}}_{\hat 0}&=&+ \, {\omega^{\hat 0}}_{\hat 1}=\Phi^{\prime}e^{-\Lambda}e^{\hat 0} \;, \nonumber \\
 {\omega^{\hat 1}}_{\hat 2}&=&- \, {\omega^{\hat 2}}_{\hat 1}=- r^{-1}e^{-\Lambda}e^{\hat 2} \nonumber \\
{\omega^{\hat 1}}_{\hat3}&=&-{\omega^{\hat 3}}_{\hat 1}=-r^{-1}e^{-\Lambda}e^{\hat 3} \;, \nonumber \\
 {\omega^{\hat 2}}_{\hat 3}&=&-{\omega^{\hat 3}}_{\hat 2} = -r^{-1}\cot\theta \,e^{\hat 3} \;. 
\end{eqnarray}
Note that  the last component is (as necessary) the same as for the general affine connection $A$,
and that the nonzero components of the contorsion tensor are then found to be (modulo the antisymmetry
with respect to the first two spatial indices in the second equation)
\bea \label{contorsion}
{K^{\hat{1}}}_{\hat{0}\hat{0}}&=& {K^{\hat{0}}}_{\hat{1}\hat{0}}=V -  e^{-\Lambda}\Phi^{\prime} \,,\nonumber \\
{K^{\hat{1}}}_{\hat{2}\hat{2}}&=& {K^{\hat{1}}}_{\hat{3}\hat{3}}=W +  r^{-1}e^{-\Lambda} \,.
\eea
Because of the restricted number of nonzero components,  the nonzero components of the torsion tensor $T_{i [jk]}$
(which is antisymmetric with respect to the last two indices,  are the same (modulo some permutation of indices) as those 
of the contorsion tensor  $K_{i j k}= K_{[i j] k}$
(which is antisymmetric with respect to the first two indices), e.g.
\bea
T_{\hat{0} [\hat{1}\hat{0}]}&=& K_{\hat{0} \hat{1}\hat{0}} = -K_{\hat{1} \hat{0}\hat{0}}=- {K^{\hat{1}}}_{\hat{0}\hat{0}}\nonumber\\
T_{\hat{2} [\hat{1}\hat{2}]}&=& K_{\hat{2} \hat{1}\hat{2}} = K_{\hat{3} \hat{1}\hat{3}} = T_{\hat{3} [\hat{1}\hat{3}]}= - {K^{\hat{1}}}_{\hat{2}\hat{2}}
\eea

Using \eqref{ACompts} we can construct the Einstein tensor of the $A$ connection:
\be
G_{ij}(A) \equiv F_{ij}(A)-\frac{1}{2}\eta_{ij}\, F(A)
\ee
This tensor happens to be symmetric, $G_{ij}(A)=G_{ji}(A)$ under our (static, spherically symmetric)
assumptions.
Its nonzero components read,
\begin{eqnarray} \label{G}
G_{\hat{t}\hat{t}}&=&\frac{1}{r^2}-W^2+2e^{-\Lambda}\frac{(rW)^{\prime}}{r}   \nonumber \\
G_{\hat{r}\hat{r}}&=&-2VW - \frac{1}{r^2}+W^2 \nonumber \\
G_{\hat{\theta}\hat{\theta}}&=&G_{\hat{\phi}\hat{\phi}}=-VW - e^{-\Lambda}\frac{(rW)^{\prime}}{r} + e^{-\Phi-\Lambda}(e^{\Phi}V)^{\prime}  \nonumber \\
\end{eqnarray}
For additional clarity, we used here a more explicit notation for the frame indices:
\be
\hat{t}=\hat{0} \, , \, \hat{r}=\hat{1} \, , \,
\hat{\theta}= \hat{2} \,, \,\hat{\phi}= \hat{3} \, \,.
\ee

\section{Torsion bigravity action}

Using the previous formulas we can now write down the action,
and derive from it the field equations. [We have checked that varying the spherically-symmetric-reduced 
action does yield field equations that are equivalent to the spherically-symmetric-reduced field equations, as derived directly
from the general field equations in Ref. \cite{Rauch:1981tva}.]
We recall that the structure of the action is
\be
 S=S_{\rm{field}}+S_{\rm{m}} \;.
 \ee
The variation of the matter action $ S_{\rm{m}}$ with respect to the metric reads
\be
\delta S_{\rm{m}}=\int\delta (\sqrt{g} L_m) d^4x  = \frac{1}{2}\int\sqrt{g}T^{\mu\nu}\delta g_{\mu\nu}d^4x
\ee
while we assume here that its variation with respect to the SO(3,1) $A$ connection (linked to the local,
quantum, spin density) vanishes.

The field action is the sum of various contributions:
\be
S_{\rm{field}}=S_R+S_F+S_{F^2} =\int d^4x  \sqrt{g}\left\{ L_R+L_F+L_{F^2} \right\} \label{SField}\,.
\ee
Here (neglecting to write the ``double-zero" term $\propto F_{[ij]}^2$)
\bea
L_R&=&c_R R[g] \;, \nonumber \\
L_F &=& c_F F[g, A] \;, \nonumber \\
 L_{F^2}&=& c_{F^2}\left(F_{(ij)}^2 - \frac{1}{3}F^2\right)  \;,
\eea
and
\be
d^4x  \sqrt{g}=  dt \, ( w(r) \,  dr )\, (\sin \theta \, d \theta \,d\phi) \,,
\ee
where
\be
w(r) \equiv r^2e^{\Phi+\Lambda}\,.
\ee
For notational simplicity, we shall often omit below to include in the action the trivial (field-independent) volume factor
$dt \, (\sin \theta \, d \theta \,d\phi)$, so as to work with a radial action $S^{\prime}=\int dr w(r) L$.

The usual Einstein-Hilbert term is explicitly computed as being
\bea
w \, R(g)&=&r^2e^{\Phi+\Lambda}\left[  -4e^{-\Lambda}\frac{\left(e^{-\Lambda}\right)^{\prime}}{r} - 2e^{-\Phi-\Lambda}\left( e^{\Phi-\Lambda}\Phi^{\prime}  \right)^{\prime}  \right. \nonumber \\ 
&&  \left. +\frac{2}{r^2} - \frac{2e^{-2\Lambda}}{r^2}  -4\frac{e^{-2\Lambda}}{r}\Phi^{\prime}  \right] \;,
\eea
which can be rewritten in the form
\be
w \, R(g)= 2 e^{\Phi+\Lambda}\frac{d}{dr}\left[ r\left(  1-e^{-2\Lambda}  \right) \right] + \frac{d}{dr} Q(r) \;,
\ee
where
\be
Q(r)=  -2r^2e^{\Phi-\Lambda} \Phi^{\prime} \; .
\ee
Note that in this form, the first term is linear in the first derivatives of the metric variables (actually linear in $\Lambda^{\prime}$).
The affine-connection analog of the Einstein-Hilbert term is obtained by inserting Eqs. \eqref{G} in
\be
w \, F(A)= - w \, G(A)= w \, \left[ G_{\hat{t}\hat{t}}- G_{\hat{r}\hat{r}} - 2 \, G_{\hat{\theta}\hat{\theta}}\right]
\ee
where we used the fact that
\be
 G(A) =  \eta^{ij} \left(F_{ij}(A)-\frac{1}{2}\eta_{ij}F(A) \right) = - F(A)\,.
\ee
To streamline the structure of the terms depending on the derivatives of $V$ and $W$, it is useful to introduce a shorthand
notation for the kind of covariant derivatives of $V$ and $W$ entering Eqs. \eqref{G}, namely
\begin{eqnarray}
\nabla V&\equiv& e^{-\Phi-\Lambda}(e^{\Phi}V)^{\prime}=e^{-\Lambda}\left(V^{\prime} + \Phi^{\prime} V\right) \\
\nabla W &\equiv& e^{-\Lambda}\frac{(rW)^{\prime}}{r}= e^{-\Lambda}\left( W^{\prime} + \frac{W}{r}\right) \;.
\end{eqnarray}
We also introduce a shorthand notation for the term involving the square of $W$, namely
\be
W^2_{-} \equiv W^2 - \frac1{r^2} \,.
\ee
With this notation, we have
\be
F(A)= 4\nabla W - 2\nabla V + 4V W - 2W^2_{-}  \; \; .
\ee

Concerning the contribution quadratic in $F_{ij}(A)$, it is easy to see that 
\be
 F_{ij}^2 = \left (F_{ij}-\frac{1}{2}\eta_{ij}F\right)^2 = G_{ij}^2
\ee
so that $ L_{F^2}$ can be directly expressed in terms of $G_{ij}(A)$ as
\be
 L_{F^2}= c_{F^2}\left(G_{(ij)}^2 - \frac{1}{3}G^2\right)  \;.
\ee
Inserting the expressions \eqref{G} for the components of $G_{ij}$, and using the shorthand notation introduced above,
leads to
\bea
&& \frac32 \left(G_{(ij)}^2 - \frac{1}{3}G^2\right) =  (\nabla V+ \nabla W)^2  + 2\nabla V(VW-2W^2_{-}) \nonumber \\
 && + 2 \nabla W(-5VW + W^2_{-})+(VW+W^2_{-})^2 
\eea
At this stage, the various contributions to the action take the form
\begin{eqnarray}
w \,L_R&=&2c_R e^{\Phi+\Lambda}\frac{d}{dr} \left( r(1-e^{-2\Lambda}) \right) + \frac{d}{dr}\left(c_R Q(r) \right)\,,\nonumber \\
w \,L_F&= &c_F r^2 e^{\Phi+\Lambda}(4\nabla W - 2\nabla V + 4V W - 2W^2_{-}) \,, \nonumber \\
w \,L_{F^2}&= &\frac{2}{3}c_{F^2}r^2e^{\Phi+\Lambda}\left\{  (\nabla V+ \nabla W)^2  \right.\nonumber \\
&& + 2\nabla V(VW-2W^2_{-}) +2 \nabla W(-5VW + W^2_{-}) \nonumber \\
& &\left. +(VW+W^2_{-})^2\right\}  \;.
\end{eqnarray}
A remarkable fact about this action is that the only term containing the square of derivatives is
the contribution $\propto  (\nabla V+ \nabla W)^2$ in $L_{F^2}$. It is then convenient to add a 
  so-called "double-zero" term to the action, so as to end up with an equivalent action  which is only {\it linear in derivatives}. 
[In the present case, this is also equivalent to making a Legendre transform.]

To explain the idea behind this transformation, let us first consider a toy model with the Lagrangian
\be
L^{\rm{old}}_{\rm{toy}}=\dot{q}^2+2A(q)\dot{q}-V(q) \;.
\ee
We can eliminate the square of the derivative of $q$ by adding  the following double-zero term to the Lagrangian,
involving a new, independent variable $\pi$:
\be
\Delta L(\pi, \dot{q}, q)=-\left[ \pi-(\dot{q}+A(q))  \right]^2 \;.
\ee
Indeed, the equation of motion of $\pi$ obtained by varying $L^{\rm{old}}_{\rm{toy}}+  \Delta L$ is
\be
-2 \left[\pi - (\dot{q}+A(q)) \right]=0 \;.
\ee
Then the modification of the equation of motion of $q$ coming from varying $\Delta L(\pi, \dot{q}, q)$
will involve (because of the quadratic nature of $\Delta L(\pi, \dot{q}, q)$) a factor $\left[\pi - (\dot{q}+A(q)) \right]$,
which vanishes when $\pi$ is on-shell. This shows that the action 
\be
L^{\rm{new}}_{\rm{toy}}(\pi, \dot{q}, q)=L^{\rm{old}}_{\rm{toy}}(\dot{q}, q) + \Delta L(\pi, \dot{q}, q)
\ee
leads to equivalent equations of motion. But the latter action is first-order in derivatives. Indeed:
\bea
&& L^{\rm{new}}_{\rm{toy}}(\pi, \dot{q}, q)=L^{\rm{old}}_{\rm{toy}}-\left[ \pi-(\dot{q}+A(q))  \right]^2 \nonumber\\
&=&  2\pi[\dot{q}+A(q)] - \pi^2-A(q)^2  -V(q)  \nonumber \\ 
&=& 2 \pi \dot{q} - (\pi-A(q))^2  -V(q) 
\eea
On the last line we recognize the result of making a Legendre transformation from $\dot{q}$
to $2 \pi = \delta L^{\rm{old}}_{\rm{toy}}/\delta \dot{q}$.

In our case, we choose to introduce as new variable the only combination of covariant derivatives of $V$ and $W$
that enters quadratically in the action, namely
\be
\pi=\nabla V + \nabla W \;.
\ee
We then add to the original action the double-zero term
$$
-  \frac{2}{3}c_{F^2}r^2e^{\Phi+\Lambda} \left(\nabla V+ \nabla W -\pi \right)^2
$$
which yields
\begin{eqnarray}
w \,L_{F^2}^{\rm{new}} &=&  \frac{2}{3}c_{F^2}r^2e^{\Phi+\Lambda}\left\{ 2\pi (\nabla V+ \nabla W) \right.\nonumber \\
& & \left.  - \pi^2 + 2\nabla V(VW-2W^2_{-})  \right. \nonumber \\
 & & \left. +2 \nabla W(-5VW + W^2_{-})+(VW+W^2_{-})^2\right\}  \;. \nonumber\\
\end{eqnarray}
The field action then looks as follows,
\begin{eqnarray}  \label{Lpi}
S^{\prime}_{\rm{field}}&=& \int dr \left\{ 2c_R e^{\Phi+\Lambda}\d_r\left[ r(1-e^{-2\Lambda}) \right] \right.\nonumber \\
& & + c_F r^2 e^{\Phi+\Lambda}(4\nabla W - 2\nabla V + 4V W - 2W^2_{-}) \nonumber \\
& &+\frac{2}{3}c_{F^2}r^2e^{\Phi+\Lambda}\left[ 2\pi (\nabla V+ \nabla W) - \pi^2 \right.\nonumber \\
&& + 2\nabla V(VW-2W^2_{-}) +2 \nabla W(-5VW + W^2_{-}) \nonumber \\
& & \left. \left. +(VW+W^2_{-})^2\right] \right\}   \;.
\end{eqnarray}

We use the macroscopic energy-momentum tensor,
\be
T^{\mu\nu}=\left[ \rho(r)+P(r)\right]u^{\mu}u^{\nu}+P(r)g^{\mu\nu} \;,
\ee
i.e., using $u^0= e^{-\Phi}$,
\begin{eqnarray} \label{MacroEMT}
T^{00}&=&\left[ \rho(r)+P(r)\right] e^{-2\Phi}-P(r)e^{-2\Phi} = \rho(r)e^{-2\Phi} \nonumber \\
T^{rr}&=&P(r)g^{rr}=P(r)e^{-2\Lambda} \nonumber \\
T^{\theta\theta}&=&P(r)g^{\theta\theta}=\frac{P(r)}{r^2} 
\end{eqnarray}
so that the variation of the matter action reads
\bea
\delta S^{\prime}_m &=& \frac{1}{2}\int dr \, w \left[   \rho(r) e^{-2\Phi} \delta (-e^{2\Phi}) + P(r) e^{-2\Lambda}\delta(e^{2\Lambda})  \right] \nonumber \\
&=& \int dr \,r^2 e^{\Phi + \Lambda} [ - \rho(r)\delta\Phi + P(r)\delta \Lambda ]\,.
\eea

\section{ Torsion bigravity field equations}

Let us now write down the equations obtained from varying the action $S^{\prime}_{\rm{field}}+ S^{\prime}_m=\int dr w(r) L$
(considered in its first-order form, with $\pi$ as an independent variable) with respect to the five field variables $x^a= (\Phi, \Lambda, V, W, \pi)$, $a=1, 2, 3, 4, 5$. Note that, introducing $x^0\equiv r$, as a fictitious sixth timelike variable, the latter first-order action has the structure 
\be \label{CartanForm}
S^{\prime}=\int dx^0 \left[ A_a(x) {\dot x}^a + A_0 \right] = \int A_\mu(x) dx^\mu
\ee
Here, we denoted ${\dot x}^a = dx^a/dx^0$, and $x^\mu=(x^0, x^a)$, with $\mu=0,1, 2, 3, 4, 5$. The six components
$A_\mu(x^\nu)$ of the one-form $A_\mu(x) dx^\mu$ depend on the six variables $x^\nu$. The one-form $A_\mu(x) dx^\mu$
is just the usual Hamilton-Cartan one-form $p_a dq^a - H dt$ of a first-order action, but we find useful to view it as the Maxwelllike
action for a {\it massless} charged particle of worldline $x^\mu$ interacting with an external electromagneticlike potential $A_\mu(x^\nu) $.

Let us write  separately the contributions coming from varying the various pieces of the action $S^{\prime}=\int dr w(r) L$
with respect to the five field variables $x^a= (\Phi, \Lambda, V, W, \pi)$.

\begin{widetext}
\begin{eqnarray}
 \frac{\delta ( w \,L_R)}{\delta \Phi}=& & 2c_R e^{\Phi+\Lambda}\d_r\left[ r(1-e^{-2\Lambda}) \right] \label{Term1} \\
\frac{\delta ( w \,L_R)}{\delta \Lambda}=& & 2c_R e^{\Phi+\Lambda}\left[ 1-e^{-2\Lambda}(1+2rF) \right] \\
\frac{\delta ( w \,L_F)}{\delta \Phi}=& &  c_F \left[  r^2e^{\Phi+\Lambda}\left( 4\nabla W+4VW - 2W^2_{-} \right) +4re^{\Phi}V  \right]\\
\frac{\delta ( w \,L_F)}{\delta \Lambda}=& & c_F r^2e^{\Phi+\Lambda}\left( 4VW-2W^2_{-} \right) \\
\frac{\delta ( w \,L_F)}{\delta V}=& & 4c_F r^2e^{\Phi+\Lambda}W + 4 c_F re^{\Phi}\\
\frac{\delta ( w \, L_F)}{\delta W}=& & -4c_F r(re^{\Phi})^{\prime} + 4 c_F r^2e^{\Phi+\Lambda}(V-W)  \\
\frac{\delta ( w \,L_{F^2})}{\delta \Phi}=& &  \frac{2}{3}c_{F^2}\left\{ -e^{\Phi}V\left( 2r^2\pi \right)^{\prime} +r^2e^{\Phi+\Lambda}\left[ 2\pi\nabla W - \pi^2 + 2\nabla W (-5VW+W^2_{-}) +(VW+W^2_{-})^2 \right] \right. \nonumber \\
& & \left. - e^{\Phi}V\left[ 2r^2(VW-2W^2_{-}) \right]^{\prime}  \right \} \\
\frac{\delta ( w \,L_{F^2})}{\delta \Lambda}=& &  \frac{2}{3}c_{F^2}r^2e^{\Phi+\Lambda}\left[ -\pi^2 + (VW+W^2_{-})^2 \right] \\
\frac{\delta ( w \,L_{F^2})}{\delta V}=& & \frac{2}{3}c_{F^2}\left[  -e^{\Phi}(2r^2\pi)^{\prime} + 2r^2W(e^{\Phi}V)^{\prime} - e^{\Phi}(2r^2VW)^{\prime} + e^{\Phi}(4W^2_{-}r^2)^{\prime} \right. \nonumber \\
& & \left. - 10r^2e^{\Phi+\Lambda}\nabla W W + 2r^2e^{\Phi+\Lambda}(VW+W^2_{-})W  \right] \\
\frac{\delta ( w \,L_{F^2})}{\delta W}=& & \frac{2}{3}c_{F^2}\left[  -r(2re^{\Phi}\pi)^{\prime} + 2r^2e^{\Phi+\Lambda}\nabla V \;V - 8r^2e^{\Phi+\Lambda}\nabla V\;W  - 10re^{\Phi}V(rW)^{\prime}  \right. \nonumber \\
& & \left.+ 10r(re^{\Phi}VW)^{\prime} + 4re^{\Phi}W(rW)^{\prime} - 2r(re^{\Phi}W^2_{-})^{\prime} + 2r^2e^{\Phi+\Lambda}(VW+W^2_{-})(V+2W) \right] \\
\frac{\delta ( w \,L_{F^2})}{\delta \pi}=& & \frac{4}{3}c_{F^2}r^2e^{\Phi+\Lambda}(\nabla V + \nabla W - \pi)  \\
\frac{\delta ( w \,L_m)}{\delta\Phi}&=&-r^2e^{\Phi+\Lambda} \rho(r)  \\
\frac{\delta ( w \,L_m)}{\delta\Lambda}&=&r^2e^{\Phi+\Lambda}P(r) \label{TermLast}
\end{eqnarray}

\end{widetext}
Here we have introduced (after variation) the shorthand notation $F$ for the radial derivative of $\Phi$:
\be
F \equiv \Phi^{\prime} \;.
\ee

We use as basic equations for the five field variables $x^a= (\Phi, \Lambda, V, W, \pi)$ the  five first-order equations 
\be
E_a\left(\frac{d x^b}{d r}, x^c, r \right)=0  \; , \; a=1, 2, 3, 4, 5
\ee
with (denoting $ c\equiv c_F$, so that $ \lam - c \equiv c_R$)
\begin{eqnarray}
&E_1\equiv &-\frac{3 \k^2}{2}r^2(c-\lam)e^{\Lambda-\Phi} \,\frac{\delta ( w\,L)}{\delta \Lambda} \label{E1} \\
&E_2\equiv&-\frac{3 \k^2}{2}r^2(c-\lam)e^{\Lambda-\Phi} \,\frac{\delta ( w\,L)}{\delta \Phi} \\
&E_3\equiv& -\frac{3 \k^2}{2c}r(c-\lam)e^{-\Phi}  \,\left(\frac{\delta ( w\,L)}{\delta V}- \frac{\delta ( w\,L)}{\delta W}\right) \\
&E_4\equiv& \frac{3\k^2}{4c}r(c-\lam)e^{-\Phi}  \,\frac{\delta ( w\,L)}{\delta W} \\
&E_5\equiv&\frac{3\k^2(c-\lam)}{4cr\lam}e^{-\Phi} \,\frac{\delta ( w\,L)}{\delta \pi} \label{E5}
\end{eqnarray}
where each term $\delta ( w\,L)/\delta x^a$ is obtained by summing the corresponding terms among 
Eqs. \eqref{Term1}-\eqref{TermLast}. [The factors $\k^2 (c-\lam)= - \k^2 c_R$ have been included to eliminate
the denominator implicitly present in $c_{F^2}= \frac{\eta \lam}{\k^2}= \frac{c \lam}{(\lam-c) \k^2}$.]

The five (geometric) field equations above must be supplemented (when considering the interior of a star)
by the usual (universal) matter equation following from the (radial) conservation law $\nabla^{g}_\mu T^{\mu \nu}=0$
for a spherically-symmetric configuration 
with macroscopic energy-momentum tensor, Eqs. \eqref{MacroEMT}, namely
\be
E_m=0 \,,
\ee
with
\be \label{MatterEq1}
E_m \equiv  P^{\prime} + (\rho+P) \frac{d\Phi}{dr} \equiv  P^{\prime} + (\rho+P) F
\ee

\section{Reduction of the field equations to a ghost-freelike system of three first order equations}\label{NonlinSys}

Let us recall that the basic aim of the present work is to study the geometric torsion bigravity model as an
alternative to the usually considered bimetric gravity models. The latter models are defined by considering
two independent dynamical metric tensors, say $g_{\mu \nu}$ and $f_{\mu \nu}$, having separate Einstein-Hilbert
actions, and being coupled to each other (besides some matter coupling)  via some generalized Fierz-Pauli potential
${\cal V}(f,g)$. These models are generalizations of the massive gravity models where the metric $f_{\mu \nu}$
is non dynamical, and frozen into some given background value (e.g. a Minkowski background $f_{\mu \nu}= \eta_{\mu \nu}$).
For many years, it was thought that massive gravity models (and, consequently, their bimetric generalizations)
were plagued by the necessary presence of an additional, ghostlike, degree of freedom 
\cite{Boulware:1973my,Deffayet:2005ys,Creminelli:2005qk}.
The latter Boulware-Deser ghost enters only at the nonlinear level (because, at the linear level, the Fierz-Pauli potential
\cite{Fierz:1939zz,Pauli:1939xp,Fierz:1939ix} ensures the presence of only five, healthy degrees of freedom
in the massive-gravity sector).

It was emphasized by Babichev, Deffayet and Ziour \cite{Babichev:2009us} that the presence of the Boulware-Deser ghost
in generic massive gravity models\footnote{We start by considering generic massive-gravity (and bimetric)
models containing a Boulware-Deser ghost to contrast them with the properties of ghost-free massive-gravity (and bimetric) 
models.} is already apparent when considering (co-diagonal) spherically symmetric solutions. More precisely,
a generic massive gravity model has (when using a Schwarzschild radial coordinate $r$ for the physical metric $g_{\mu \nu}$) 
three variables: $\Phi(r)$, $\Lambda(r)$ (defined as in Eq. \eqref{ds2} above) together with a third ``gauge" variable
$\mu(r)$ relating the Schwarzschildlike radius $r$ to the ``flat" radial variable $r_f$ defined by the background metric $f_{\mu \nu}$, namely $r_f= r e^{- \mu(r)/2}$.
The crucial point (which can also be seen in the explicit  field equations of Ref. \cite{Damour:2002gp})
is that the massive-gravity field equations are first order in $\Phi(r)$, and $\Lambda(r)$, but {\it second order} in $\mu(r)$.
This means that the total differential order of the massive-gravity $\Phi(r)$, $\Lambda(r)$, $\mu(r)$ system 
is {\it four}. Equivalently,
the general\footnote{Here, ``general" means that we do not impose boundary conditions at infinity.} 
exterior spherically-symmetric solution of a generic massive-gravity model contains four arbitrary integration
constants. One of them will be an additional constant $c_0$ in $\Phi(r)$, which is physically irrelevant because it can be
gauged away by renormalizing the time variable: $t \to t'= e^{- c_0/2} t$. We conclude that 
the general exterior spherically-symmetric solution of a generic (ghostfull) massive-gravity model contains {\it three} 
physically relevant arbitrary integration constants. This is {\it one more constant} than for the general 
exterior spherically-symmetric solution of the Fierz-Pauli {\it linearized massive gravity} model. 
Indeed, the latter general linearized solution for $h_{\mu \nu}=g_{\mu \nu} - f_{\mu \nu}$ is (see \cite{Boulware:1973my})
\bea \label{hmunuBD}
h_{00}&=&2 \, Y_\k(r) \nonumber \\
h_{0i}&=& 0 \nonumber \\
h_{ij}&=& \delta_{ij} Y_\k(r) - \frac{1}{\kappa^2} \d_i \d_j Y_\k(r)
\eea
where $\k$ denotes the mass of the massive graviton, and where $Y_\k(r)$ is the general exterior spherically-symmetric solution
of the Yukawa equation 
\be
(\Delta - \k^2)Y_\k =\frac{16 \pi G_\k T^\mu_\mu}{3} \,,
\ee
 which contains {\it two} integration constants, $c_+, c_-$, namely
\be
Y_\k(r)= c_+ \frac{e^{+\k r}}{r} +  c_- \frac{e^{-\k r}}{r}\,.
\ee
We recall in passing that the trace of  $h_{\mu \nu}$ is locally related to the matter density via
\be\label{hBD}
h^\mu_\mu= - \frac{16 \pi G_\k T^\mu_\mu}{3 \,\k^2}\,.
\ee
Summarizing:  the presence of a sixth field degree of freedom in a generic (ghostfull) massive-gravity model is visible
when considering the general exterior spherically-symmetric solution: indeed, the latter solution, generically
involves three physically relevant integration constants, which is one more than the two physically relevant integration constants
$c_+$, $c_-$ entering the corresponding linearized solution of the five-degree-of-freedom Fierz-Pauli model.
In addition, we recall that the linearized massive-gravity solution necessarily involves a $\frac1{\k^2}$ factor in 
some of its components, and that this feature is the origin of the appearance of a Vainshtein radius below which
one cannot trust the usual weak-field perturbation expansion of massive gravity \cite{Vainshtein:1972sx,Damour:2002gp}.

Let us emphasize that the ability of the spherically-symmetric limit to detect the presence of the Boulware-Deser ghost
is somewhat obscured if one focusses, from the beginning, on exponentially decaying solutions, rather than on
general exterior solutions. [See, in this respect,
Refs. \cite{Damour:2002gp}, \cite{Babichev:2009us} and \cite{Babichev:2010jd}.]

When extending a massive-gravity model into a corresponding bimetric gravity one, we must add to the count of the
physically relevant integration constants entering a general exterior solution the Schwarzschildlike mass $m$ parametrizing
the physics of the massless- spin-2 sector. We therefore conclude that the general
exterior solution of a ghostfull bimetric gravity model will involve {\it four} physically relevant integration constants, 
while the general exterior solution of a ghost-free bimetric gravity model will involve only {\it three} physically relevant 
 integration constants (corresponding to $m$, $c_+$, $c_-$ parametrizing the corresponding linearized system).
 [We recall that we discounted here the physically irrelevant additional constant entering $\Phi(r)$.]
 The fact that the general exterior solution of  ghost-free bimetric gravity models (using the restricted class
 of potential ${\cal V}(f,g)$ discovered in \cite{deRham:2010kj}) indeed involves only {\it three} 
 physically relevant integration constants has been explicitly shown by Volkov \cite{Volkov:2012wp}. 
 Indeed, he showed how to reduce the (co-diagonal) field equations to a system of {\it three first-order} differential
 equations, for the three variables $N$, $Y$, and $U$, see Eqs. (5.7) in \cite{Volkov:2012wp}. [The variable $\Phi= \ln Q$
 is then obtained by a quadrature: $\Phi = \int dr {\cal F}_5 + c_0$, see Eq. (5.3c) in \cite{Volkov:2012wp}.]
 
 We are now going to show that the torsion bigravity model is {\it similar to the ghost-free  bimetric gravity} models
 in that its general exterior spherically-symmetric solution only involves {\it three} physically relevant 
 integration constants. [We will see later that these three integration constants do
 correspond  to the constants $m$, $C_+$, $C_-$ parametrizing the corresponding linearized 
 torsion bigravity system.] This will be shown by reducing the system of five first-order field equations $E_1$--$E_5$
 written in the previous section to a system of {\it three first-order} differential equations (together with
 a quadrature for $\Phi(r)$). In view of the fact, recalled above, that the presence of the Boulware-Deser
 ghost was visible in spherically-symmetric solutions of generic ghostfull bimetric gravity models, we consider
 this property of torsion bigravity as a suggestion (though not a proof) that it might be ghost-free in a
 general (time-dependent and non-spherically-symmetric) situation.
 
 As our reduction process is algebraically involved, we will not display all the technical details,
 but only explain the algorithm by which we could explicitly derive a reduced system of three first-order
 equations for three unknowns. Explicit calculations are anyway better done by using algebraic
 manipulation programmes, starting from the explicit basic field equations written in the previous section.

 Before explaining the explicit reduction process we used, let us briefly indicate how the reduction issue
 could be formulated in terms of the Hamilton-Cartan action \eqref{CartanForm}. The variational equations
 of motion coming from the first-order action \eqref{CartanForm} are
 \be \label{Fv}
 {\cal E}_\mu \equiv F_{\mu \nu}(x) \frac{d x^\nu}{d x^0}=0 \,,
 \ee
 where $ F_{\mu \nu}(x) = \d A_{\nu}(x)/\d x^\mu - \d A_{\mu}(x)/\d x^\nu$  are the components
(with $\mu=0,1, 2, 3, 4, 5$) of the two-form $F = dA$, and
 where we recall that $x^0$ simply denotes the radial variable $r$, which plays the role of time in our action.
 Because of the antisymmetry of $ F_{\mu \nu}$, there are only five independent  equations among 
 the equations ${\cal E}_\mu$, Eq. \eqref{Fv} (say $ {\cal E}_a$, for $a=1,2,3,4,5$).
 A necessary condition for the variational equations \eqref{Fv} to have nontrivial solutions in the phase-space ``velocity" 
 $v^\mu = \frac{d x^\nu}{d x^0}$ is that the determinant of the six-by-six matrix $ F_{\mu \nu}$ be vanishing. 
 As $ F_{\mu \nu}$ is antisymmetric and even, its determinant is the square of its Pfaffian 
 \be
 {\rm Pf}[F] \equiv \epsilon^{\mu_1 \nu_1 \mu_2 \nu_2 \mu_3 \nu_3} F_{\mu_1 \nu_1} F_{\mu_2 \nu_2} F_{\mu_3 \nu_3}\,.
 \ee
 This shows that a necessary condition following following from the five equations $ {\cal E}_a =0$ (which are
 equivalent to the equations $E_a=0$ of the previous section) is the constraint
 \be \label{Pf=0}
 {\rm Pf}[F(x)]=0\,.
 \ee
 The latter  constraint is purely algebraic in the five variables  $x^a= (\Phi, \Lambda, V, W, \pi)$ 
 (and depends on $x^0=r$). In turn, the (primary) constraint \eqref{Pf=0} implies as secondary constraint an
 equation linear in the velocities $v^a = \frac{d x^a}{d x^0}$ (i.e. the radial derivatives of $(\Phi, \Lambda, V, W, \pi)$),
 namely
 \be \label{dPf=0}
 0= \frac{d {\rm Pf}[F(x)]}{d x^0}=  \frac{d x^\mu}{d x^0} \frac{\d {\rm Pf}[F(x)]}{\d x^\mu}\,.
 \ee
This argument indicates that the basic system of five equations $E_1$--$E_5$ of the previous section implies (at least)
one algebraic constraint, \eqref{Pf=0}, together with the extra differential condition \eqref{dPf=0}. To check what is
the precise import of these constraints on the number of free data determining the general exterior solution of our system
we need to explicitly write down and study these constraints, as we will do next (starting directly from the explicit
form \eqref{E1}--\eqref{E5} of our five basic equations $E_1$--$E_5$).

When doing so, it it convenient to start by noticing that the gauge symmetry $t \to t'= e^{- c_0/2} t$,
which corresponds to changing $\Phi(r)$ into $\Phi(r) + c_0$ shows that our basic five field equations
can be entirely expressed in terms of $F(r) \equiv \Phi^{\prime} $, without any explicit appearance of
the undifferentiated variable $\Phi(r)$. Actually, the various $e^{-\Phi}$ factors in our definitions 
\eqref{E1}--\eqref{E5} were designed to realize this disappearance of $\Phi(r)$. In other words, we
can consider the system $E_1$--$E_5$ as being algebraic in $F$, and differential (of first order)
only in the four variables $\Lambda$, $V$, $W$, and $\pi$. 

It is also useful to work with a slightly modified set of variables. In the following we shall replace the set of variables
$F$, $\Lambda$, $V$, $W$, and $\pi$ by the new set $F$, $L$, $V$, $\bY$, and $\pi$ where
\bea \label{defs}
L &\equiv&  e^{\Lambda} \,,\\
\bY &\equiv& Y + \frac{1}{r} \equiv V + W + \frac{1}{r} \,,
\eea
where we used also the intermediate notation
\be
 Y \equiv V + W\,.
\ee
The usefulness of this change of variables is that it allows one to easily show that two combinations
of our five basic equations  $E_1$--$E_5$, \eqref{E1}--\eqref{E5}, yield {\it two algebraic constraints}
in the five variables $F$, $L$, $V$, $\bY$, $\pi$.

On the one hand, the equation $E_1$ turns out to be algebraic in $F$, $L$, $V$, $\bY$, $\pi$ (without involving
any derivative):
\be
E_1=E_1(F, L, V, \bY, \pi; P)\;.
\ee
Moreover $E_1$ is {\it linear} in $F$ and quadratic in $L$. 
As indicated, $E_1$ also involves the pressure $P(r)$ as a matter source.
The constraint $E_1 =0$ will be used to algebraically eliminate  $F$ by expressing it in terms of the other variables. 

On the other hand, the only derivative entering the two equations $E_3$ and $E_5$ is $ \overline{Y}^{\prime}$.
This implies that a linear combination of $E_3$ and $E_5$ yields an algebraic constraint. More precisely the new expression
\be
E_{35} \equiv  rE_3  -2 r^3 \lam Y E_5 
\ee
is an algebraic expression in our (redefined) variables, namely
\be \label{E35}
E_{35} = E_{35}(F, L, V, \bY, \pi)
\ee
which is  is {\it linear} in both $F$ and $L$.

The reduction process we use is then the following. First, we solve the algebraic constraint $E_1(F, L, V, \bY, \pi)=0$
(which is linear in $F$) with respect to $F$ to get
\be \label{Fsol}
F = F_{\rm sol}[L, V, \bY, \pi; P]\,.
\ee
Then, we replace $F \to F_{\rm sol}[L, V, \bY, \pi; P]$ in the other algebraic constraint $E_{35}$, Eq. \eqref{E35}, to get
a reduced algebraic constraint involving only the four geometric variables $L, V, \bY, \pi$, say
\be
E^{\rm red}_{35}(L, V, \bY, \pi; P) \equiv  E_{35}(F_{\rm sol}[L, V, \bY, \pi; P], L, V, \bY, \pi) \,.
\ee
The so-obtained algebraic constraint $E^{\rm red}_{35}(L, V, \bY, \pi)=0$ turns out to be {\it quadratic} in $L$.
There is a {\it unique root} of this quadratic equation in $L$ \footnote{This is the smallest root, i.e. the root with a negative coefficient
in front of the discriminant when writing the equation with a positive coefficient for $L^2$.}, say
\be \label{Lsol}
L =  L^-_{\rm sol}[ V, \bY, \pi;P]
\ee
 which is such that it has the physically desirable feature of asymptotically behaving 
like its Schwarzschild counterpart 
\be
L_S(r)= e^{\Lambda_S(r)}= \frac{1}{\sqrt{1- 2m_S/r}} \to 1 \; \; {\rm as} \; \; r\to +\infty
\ee
when the arguments $V, \bY, \pi$ asymptotically decay  at infinity in a  Schwarzschildlike manner.
This requirement follows from the physical requirement that the contorsion tensor 
(being entirely generated, at the linear level,
via a massive-spin-2 excitation; see below) must decay $\propto e^{- \k r}$ so that $V$ and $W$, and the corresponding
$\bY, \pi$, must asymptotically decay as their Schwarzschild counterparts, i.e. as the corresponding frame
components of the Levi-Civita connection, see \eqref{OmCompts}.

Then, by substituting $ L \to L^-_{\rm sol}[ V, \bY, \pi; P]$, from  Eq. \eqref{Lsol}, into the expression \eqref{Fsol},
we get an explicit expression for $F$ in terms of the three geometric variables $ V, \bY, \pi$, say
\be \label{Fsol}
F = F^{\rm red}_{\rm sol}[V, \bY, \pi; P] \equiv  F_{\rm sol}[L^-_{\rm sol}[ V, \bY, \pi; P], V, \bY, \pi;P]\,.
\ee
The final stage of our reduction process consists in replacing $F \to F^{\rm red}_{\rm sol}[V, \bY, \pi] $
and $L \to L^-_{\rm sol}[ V, \bY, \pi]$ into the remaining equations $E_2$, $E_4$, and $E_5$ to get
three first-order equations for the three unknowns $V$, $ \overline{Y}$, and $\pi$ (involving also 
$P$ and $\rho$ as source terms), say
\bea \label{sys30}
0 &=& E_2^{\rm red}[ V^{\prime}, \bY^{\prime}, \pi^{\prime}, P^{\prime}, V, \bY, \pi; \rho, P] \nonumber \\
0 &=&  E_4^{\rm red}[ V^{\prime}, \pi^{\prime},V, \bY, \pi; P] \nonumber \\
0 &=&  E_5^{\rm red}[ \bY^{\prime}, V, \bY, \pi; P]
\eea
By construction, these three equations are linear in all the radial derivatives.  When replacing the radial derivative of the
pressure which appears in $ E_2^{\rm red}$ by the matter equation \eqref{Peq} discussed next,
one can  solve the three equations \eqref{sys30} for the three derivatives $ V^{\prime}, \bY^{\prime}, \pi^{\prime}$
so as to get an explicit first-order radial-evolution system, say
\bea \label{sys3}
 V^{\prime} &=& DV[V, \bY, \pi; \rho, P] \nonumber \\
 \bY^{\prime}&=& D\bY[V, \bY, \pi; \rho, P] \nonumber \\
 \pi^{\prime}&=& D\pi[V, \bY, \pi; \rho, P] \,.
\eea
When considering the solution inside a star one must augment this system by the reduction of the 
equation \eqref{MatterEq1} constraining the radial evolution of the pressure, namely
\be \label{Peq}
 P^{\prime}=- (\rho+P)  F^{\rm red}_{\rm sol}[V, \bY, \pi; P] \,
\ee
and by giving an equation of state relating $\rho$ to $P$, say $\rho=\rho(P)$.

After integrating the system \eqref{sys3}, \eqref{Peq}, for the four variables $V, \bY, \pi, P$,
one can compute the values of the variables $F$, $L$ (or $\Lambda$), and $W$ by using
Eqs. \eqref{Fsol}, \eqref{Lsol}, and \eqref{defs}. Finally, the value of the gravitational potential $\Phi(r)$ is
obtained by a quadrature:
\be \label{Phi}
\Phi(r)= -\int_r^{\infty} dr' F(r')\,
\ee
where we fixed the arbitrary additional constant in $\Phi$ by the requirement that $\Phi(r) \to 0$
at radial infinity.

\section{Linearized approximation}\label{Linear}

Let us study the linearized approximation to our five basic field equations  $E_1$--$E_5$,\eqref{E1}--\eqref{E5}.
We are, in particular, interested in understanding how the linearized solutions behave in the small-mass limit 
$\kappa \rightarrow 0$. In the next section we will then consider the second-order (post-linear) solutions.
We will see that, both at the linear level, and at the postlinear level, the limit $\kappa \rightarrow 0$ of
torsion bigravity is much better behaved than in massive gravity and bimetric gravity.
Some aspects of the linearized approximation of  dynamical-torsion models have already been
considered in Refs. \cite{Hayashi:1980ir,Nikiforova:2009qr}, and in Ref. \cite{Zhang:1982jn} for
spherically symmetric solution, but our treatment will be more extensive and detailed.

In absence of material source (i.e. when $\rho \to 0$ and $P\to 0$), the torsion bigravity field equations admit the  solution
$\Phi=0$, $F\equiv \Phi^{\prime}=0$, $\Lambda=0$, $V=0$, $W=-\frac1r$, $\bY\equiv V+W+\frac1r =0$. We denote with a
subscript 1 a first-order deviation from this trivial solution, i.e. $F_1$, $\Lambda_1$, $V_1$ and $\bY_1$.
The explicit form of  the linearized approximation of the field equations  looks as follows,
\begin{widetext}
\begin{eqnarray} 
\frac{\delta S}{\delta \Lambda}: \quad {\hat E}^{\rm{lin}}_1&\equiv & 2 c r V_1 -  c r  \overline{Y_1} +  \Lambda_1 (c - \lam) - 
  F_1 r(c  - \lam) -\frac{r^2}{4}P = 0  \label{dLdLam} \\
 \frac{\delta S}{\delta \Phi}: \quad  {\hat E}^{\rm{lin}}_2 &\equiv &  c V_1^{\prime} r^2 - c  \overline{Y_1}^{\prime} r^2 + \Lambda_1^{\prime} r(c  -  \lam)+ \frac{\rho \, r^2}{4} + 2 c r V_1 - 2 c r  \overline{Y_1} +
 \Lambda_1 (c - \lam) =0  \label{dLdPhi} \\
 \frac{\delta S}{\delta V}- \frac{\delta S}{\delta W}: \quad  {\hat E}^{\rm{lin}}_3 &\equiv &   \overline{Y_1}^{\prime} r \lam - \pi_1 r \lam  + 6 \k^2 r (c - \lam)\Lambda_1 + (9 c \k^2 r^2 - \lam - 9 \k^2 r^2 \lam)V_1  \nonumber \\
  & & - 3 \k^2 r^2 (c - \lam) F_1 + (-6 c \k^2 r^2 + \lam+ 6 \k^2 r^2 \lam) \overline{Y_1}    = 0 \label{dLdV} \\
  \frac{\delta S}{\delta W}: \quad  {\hat E}^{\rm{lin}}_4 &\equiv &  V_1^{\prime} r \lam  + \pi_1^{\prime} r^2 \lam + \pi_1 r \lam  +3 \k^2 r (c - \lam)\Lambda_1 +6 \k^2 r^2 (c - \lam)V_1 \nonumber \\
  & &  -3 \k^2 r^2 (c - \lam) F_1 + (-3 c \k^2 r^2 - 2 \lam + 3 \k^2 r^2 \lam) \overline{Y_1}   =0 \label{dLdY} \\
   \frac{\delta S}{\delta \pi}: \quad  {\hat E}^{\rm{lin}}_5 &\equiv &   \overline{Y_1}^{\prime} r - \pi_1 r - V_1 +  \overline{Y_1}  =0 \label{dLdpi}
\end{eqnarray}
\end{widetext}
The hat added on the ${\hat E}^{\rm{lin}}_n$'s indicate that these equations differ by a factor from the linearization of the 
corresponding equations $E_n$, as defined in Eqs. \eqref{E1}--\eqref{E5} above. In keeping with what was already the case at the nonlinear level, the first equation ${\hat E}^{\rm{lin}}_1$ is algebraic in the variables $\Lambda_1$, $F_1$, $V_1$, $ \overline{Y_1}$ and thus can be used to express $F_1$ in terms of other three, $F_1=F_1\left( \Lambda_1, V_1,  \overline{Y_1} \right)$. Furthermore, the equation ${\hat E}^{\rm{lin}}_5$ is algebraic in $\pi_1$ 
and can be used to express $\pi_1$ in terms of the variables $V_1$, $ \overline{Y_1}$, and $ \overline{Y_1}^{\prime}$. 
Henceforth, we solve ${\hat E}^{\rm{lin}}_1=0$ for $F_1$, and ${\hat E}^{\rm{lin}}_5=0$ for $\pi_1$ so as to 
eliminate 
\bea
F_1&=&F_1^{{\hat E}^{\rm{lin}}_1}\left( \Lambda_1, V_1,  \overline{Y_1}; P \right) \label{F1lin} \,, \\
\pi_1&=& \pi_1^{{\hat E}^{\rm{lin}}_5}\left( V_1,  \overline{Y_1}, \overline{Y_1}^{\prime} \label{pi1}\right)\,,
\eea
 from the system.
 
 It is then easily seen that the replacement of Eq. \eqref{pi1} in Eq. \eqref{dLdV} eliminates the derivative 
 $\overline{Y_1}^{\prime} $ and yields an equation which is algebraic in $\Lambda_1$, $V_1$, $ \overline{Y_1}$.
 We can then use the latter algebraic equation (which is equivalent to the combination 
 ${\hat E}^{\rm{lin}}_{35} \equiv {\hat E}^{\rm{lin}}_3 - \lam \,{\hat E}^{\rm{lin}}_5$)
 to express $\Lambda_1$ in terms of $V_1$ and $ \overline{Y_1}$, say
 \be \label{L1}
 \Lambda_1 = {\Lambda_1}^{{\hat E}^{\rm{lin}}_{1} \cup {\hat E}^{\rm{lin}}_{35}}\left(V_1,\overline{Y_1}; P\right)\,.
 \ee

After inserting all the replacements Eqs. \eqref{F1lin}, \eqref{pi1}, \eqref{L1}, one ends up with two remaining equations to solve:
Eq. \eqref{dLdY}, which is second order in $ \overline{Y_1}$ and first-order in $V_1$, and Eq. \eqref{dLdPhi}, 
which is first order in $ \overline{Y_1}$, and $V_1$. The explicit form of the latter two equations is streamlined by
 introducing the new variables, 
\begin{eqnarray}
V_{m0}&\equiv& -3V_1+2 \overline{Y_1} \label{Vm0}\\
V_{mk}&\equiv& 2V_1- \overline{Y_1}\;,  \label{Vmk}
\end{eqnarray}
We find that these variables must satisfy the following equations 
\begin{eqnarray}
V_{m0}^{\prime} &+& \frac{2}{r}V_{m0}=\frac{ \rho(r)}{4\lam} - \frac{3}{4\lam}P(r) - \frac{r}{4\lam}P^{\prime}(r) \,,\label{eqVm0} \\
V_{mk}^{\prime\prime} &+& \frac{2}{r}V_{mk}^{\prime} - \left(\frac{2}{r^2}+\k^2\right)V_{mk} \nonumber \\
&=&-\frac{\rho^{\prime}(r)}{6\lam} -\frac{\k^2 r }{4 \lam}P(r) + \frac{2}{3 \lam}P^{\prime}(r) + \frac{r}{6 \lam}P^{\prime\prime}(r)  \,.\nonumber \\
\label{eqVmk} 
\end{eqnarray}
Given a solution of these two linear equations, the full linearized solution is given 
by the inverse of Eqs. \eqref{Vm0}, \eqref{Vmk}, i.e.
\bea \label{eqY1V1}
\overline{Y_1} &=& 2 V_{m0}+ 3 V_{mk} \,,\nonumber \\
V_1 &=&  V_{m0}+ 2 V_{mk} \,,
\eea
as well as by
\bea 
\frac{1}{r}\Lambda_1 &=& V_{m0} - \eta V_{mk}+ \frac{r}{4(\lam-c)}P(r) \,,\label{eqLam1} \\
F_1 &=& V_{m0}- 2\eta V_{mk} +\frac{r}{2(\lam-c)}P(r) \,. \label{eqF1}
\eea
The total differential order of the system Eqs. \eqref{eqVm0}, \eqref{eqVmk} is {\it three}, i.e. the same order
as we found above for the full, nonlinear system.

One should note the remarkable fact that these linearized-approximation equations never involve the
inverse of the squared mass $\k^2$ of the massive spin-2 excitation. This is in sharp contrast with
the corresponding linearized massive-gravity, or bimetric gravity, equations which always involve
an inverse power of $\k^2$, see, e.g., Eqs. \eqref{hmunuBD}, \eqref{hBD}. We will see below that
the absence of inverse powers of $\k^2$ persists at the postlinear order.

Let us recall the structure of the solutions of equations of type \eqref{eqVm0} and \eqref{eqVmk}, with general 
source terms on the right-hand sides,
\begin{eqnarray}
V_{m0}^{\prime} &+& \frac{2}{r}V_{m0}=S_{m0}(r) \,,\label{eqVm0Gen} \\
V_{mk}^{\prime\prime} &+& \frac{2}{r}V_{mk}^{\prime} - \left (\frac{2}{r^2}+\k^2 \right)V_{mk}=S_{mk}(r) \,.\label{eqVmkGen} 
\end{eqnarray}
 These equations have  unique solutions that are regular at the origin and decaying at infinity. 
 They are given by the following formulas,
\begin{eqnarray}
V_{m0}(r)&=&\frac{1}{r^2}\int_{0}^{r}\hat{r}^2S_{m0}(\hat{r})d\hat{r} \\
V_{mk}(r)&=&\int_{0}^{\infty}\hat{r}^2G_{\kappa}(r,\hat{r})S_{mk}(\hat{r})d\hat{r} \,.
\end{eqnarray}
In the second equation, the Green's function $G_{\kappa}(r,\hat{r})$, satisfying the equation
\be
\left[\d^2_r+\frac{2}{r}\d_r-\left( \frac{2}{r^2}+\k^2 \right)\right]G(r,\hat{r})=\frac{1}{r^2}\delta(r-\hat{r}) \;,
\ee
is constructed as
\be
G_{\kappa}(r,\hat{r}) \equiv \frac1{W}\left[ X_{>}(r)X_{<}(\hat{r})\theta(r-\hat{r}) + X_{<}(r)X_{>}(\hat{r})\theta(\hat{r}-r) \right] \;.
\ee
where $\theta(x)$ denotes Heaviside's step function, while
\be
X_{>}(r)=\d_r\left( \frac{e^{-\kappa r}}{r} \right) \quad  \text{and} \quad X_{<}(r)=\d_r\left( \frac{\sinh(\k r)}{r} \right)
\ee
are two appropriate homogeneous solutions, incorporating the boundary conditions. Namely, $X_{>}(r)$ 
decays at infinity, while $X_{<}(r)$ is regular at $r=0$. In addition, 
\be
W\equiv r^2 \left( X_{>}^{\prime}(r) X_{<}(r)-  X_{>}(r) X_{<}^{\prime}(r) \right) = \k^3
\ee
is the appropriate (constant) Wronskian of the two solutions.

Note that $V_{m0}$ and $V_{mk}$ are  ``pure" variables corresponding to the massless and massive linear excitations respectively. We then see on Eqs. \eqref{eqY1V1}, \eqref{eqLam1}, \eqref{eqF1},
 how each metric or connection variable is some combination of these two pure variables.

Let us explicitly display the above linearized solution in the simple case where the source is a {\it constant density} star,
say 
\be \label{e0}
\rho(r) = e_0 \; .
\ee
But, first, let us note that the source terms in the linearized equations \eqref{eqVm0}, \eqref{eqVmk} have
different perturbative orders of magnitude. Indeed, we can consider that the primary source of all the variables
is the matter density $\rho(r)$, and that it defines the formal expansion parameter $\varepsilon$ of our weak-field
expansion: $\rho \equiv \varepsilon \rho_1$. Here, $\varepsilon$ is a bookkeeping device, which will be
set to one at the end. The linearized variables $\Phi_1 $, $F_1$, $V_1$, etc. are first-order in $\varepsilon$.
E.g. $\Phi= \varepsilon \Phi_1+ O(\varepsilon^2)$,  $F= \varepsilon F_1+ O(\varepsilon^2)$
(where $F_1= \Phi_1^{\prime}$), etc. On the other hand, the pressure-gradient equation
\eqref{MatterEq1} has the structure
\bea
P^{\prime}&=&- (\rho+P) F\nonumber \\
&=& -(\varepsilon \rho_1+P) \left(\varepsilon F_1 +O(\varepsilon^2)\right) \;.
\eea
The boundary condition that $P(r)$ vanishes at the surface of the star then shows that the pressure $P$ is
actually of second order in $\varepsilon$:  $P=P_2\varepsilon^2 + O(\varepsilon^3)$, with
\be
P_2^{\prime}= - \rho_1 F_1 \,.
\ee
To determine $P_2$ we must first determine the value of $F_1$ generated by  $\rho \equiv \varepsilon \rho_1$.
We shall take into account, in the next section, the second-order effects induced by the source terms involving the pressure 
$P=P_2\varepsilon^2 + O(\varepsilon^3)$ in the linearized equations \eqref{eqVm0}, \eqref{eqVmk}, \eqref{eqLam1}, \eqref{eqF1}.
In the present section, we can define the pure linearized fields $F_1$, $V_1$, etc. by neglecting all
the pressure-related source terms in the field equations \eqref{eqVm0}, \eqref{eqVmk}, \eqref{eqLam1}, \eqref{eqF1},
and by using the
constant density ansatz \eqref{e0}. This leads to the following explicit solutions of the system \eqref{eqVm0}, \eqref{eqVmk}: 
\be
V_{m0}=
\begin{cases}
 \frac{m_1r}{R_s^3} \;,\quad & r\leq R_s \\
 \frac{m_1}{r^2} \;,\quad & r\geq R_s  \label{Vm0sol}
\end{cases}
\ee

\be
 V_{mk}=
 \begin{cases}
 \frac{e^{-\k R_s}(1+\kappa R_s)}{r^2} \\
 \times \left[-\frac{2m_1}{\kappa^3 R_s^3} (\kappa r \cosh(\kappa r)-\sinh(\kappa r))\right] , & r\leq R_s \\
 \frac{e^{-\k r}(1+\kappa r)}{r^2}C_{mk} , & r\geq R_s  \label{Vmksol}
 \end{cases}
 \ee
 Here, we recall that $ \lam  \equiv 1/(16 \pi G_0)$, while $R_s$ denotes the radius of the star, and we have defined
 \begin{eqnarray} 
 m_1 &\equiv & \frac{e_0 R_s^3}{12\lam} = \frac{4 \pi G_0}{3} e_0  R_s^3   \\
  C_{mk}&=& - \frac{2}{3}m_1 \mathcal{F}(z_s)\\
  z_s &\equiv& \kappa R_s \\
  \mathcal{F}(z) &\equiv & 3\left\{z\; \cosh(z) - \sinh(z)\right\}/z^3  \label{Fz}
\end{eqnarray} 
The ``form factor" $\mathcal{F}(z)$, entering the magnitude  $C_{mk}$ of $ V_{mk}$
outside the star, has been defined so that $\mathcal{F}(z) \to 1$ when its argument $z= \k R_s \to 0$.

There are apparent factors $\propto 1/\k^3$ entering the inner solution for $ V_{mk}$. However, these factors
(which come from the Wronskian $W=\k^3$ in the Green's function) are cancelled by $O(\k^3)$ terms in the
numerators. Indeed, the Green's function itself is seen to have a finite limit as $\k \to 0$, because
\bea
\lim_{\kappa\to0}X_{>}(r)&=&\d_r\left( \frac{1}{r} \right) = -\frac1{r^2}\\
 \lim_{\kappa\to0}\frac{X_{<}(r)}{\k^3}&=&\lim_{\kappa\to0} \d_r\left( \frac{\sinh(\k r)}{\k^3 r} \right)= \frac{r}{3}
\eea
This ensures that the linearized solution has a finite limit when $\kappa \rightarrow 0$ (at a fixed value of $r$). 
In the limit $\kappa \to 0$ 
(keeping fixed both $R_s$ and $r$) one has indeed the following limit for $V_{mk}$,
\be
 V_{mk}^{\;\kappa\to0} \to
 \begin{cases}
 -\frac{2m_1r}{3R_s^3} \\
 -\frac{2m_1}{3r^2} \;.
 \end{cases}
 \ee
Let us also give the expression for $F_1$, 
\be
  F_1=
 \begin{cases}
\frac{m_1r}{R_s^3} \left[ 1 + \frac43 \eta e^{-z_s} (1 + z_s) \mathcal{F}(z) \right]\;,\quad & r\leq R_s \\
\frac{m_1}{r^2} + C_1^F(z_s) \, \frac{e^{-z}(1+ z)}{z^2} \;, \quad & r\geq R_s \label{Flin}
 \end{cases}
 \ee
 where
 \be
 z \equiv \k \, r \; ; \;  z_s \equiv \k \, R_s \,,
 \ee
 and
 \be \label{C1F}
 C_1^F(z_s)=\frac43 \eta \, m_1 \,\k^2  \mathcal{F}(z_s) \,.
 \ee
 The full (interior and exterior) solutions for the other variables are easily derived from the expressions given above.
 Let us only write down here the {\it exterior} ($ r\geq R_s$) solutions for all the variables. 
 [We recall in passing that all variables have zero background values, except for 
$W= -\frac1{r} + \varepsilon W_1 + O(\varepsilon^2)$.]
 \bea
  F_1&=&\frac{m_1}{r^2}+ 2\,\eta \,C_1 \frac{e^{-\k r}(1+\k r)}{r^2} \;, \label{Lam_1sol} \\
 \Lambda_1&=&\frac{m_1}{r}+ \eta\, C_1\frac{e^{-\k r}(1+\k r)}{r}  \;, \label{Lam_1sol} \\
 V_1 &=& \frac{m_1}{r^2}- 2 \, C_1  \frac{e^{-\k r}(1+\k r)}{r^2} \;, \\
 \bY_1 &=& \frac{2m_1}{r^2}- 3 \, C_1  \frac{e^{-\k r}(1+\k r)}{r^2} \;,\\
 W_1&=&\bY_1-V_1  =\frac{m_1}{r^2}- C_1  \frac{e^{-\k r}(1+\k r)}{r^2} \;,
 \eea
 where
 \be
 C_1 \equiv \frac{2m_1}{3} \mathcal{F}(z_s) =- C_{mk} \;.
 \ee
 It is important to display also the linearized values of the two independent components of the contorsion (and torsion), 
 as defined in Eq. \eqref{contorsion}
 \bea
 {K^{\hat{r}}}_{\hat{t}\hat{t}} &=& V -  e^{-\Lambda} F \,,\nonumber \\
 {K^{\hat{r}}}_{\hat{\theta}\hat{\theta}}&=& W +  r^{-1}e^{-\Lambda}\,.
 \eea
 They read
  \be \label{Klin}
  \left[{K^{\hat{r}}}_{\hat{t}\hat{t}}\right]_1 = V_1-F_1 =- 2 \, C_1(1+\eta)  \frac{e^{-\k r}(1+\k r)}{r^2}  \,,\nonumber \ee
  \be
  \left[{K^{\hat{r}}}_{\hat{\theta}\hat{\theta}}\right]_1 = W_1 -\frac{\Lambda_1}{r}  = - C_1(1+\eta)  \frac{e^{-\k r}(1+\k r)}{r^2}  \,.
  \ee
  Note that the (con-)torsion components are exponentially decaying. [This remains true at all orders of perturbation theory.]
  By contrast, the geometric variables  $\Phi_1, F_1, \Lambda_1, V_1, W_1$ contain
  an {\it additive mixture} of massless (power-law decaying) and massive (exponentially decaying) spin-2 excitations.

\section{Second order perturbations}

Let us consider the solutions of torsion bigravity at the second order in the source $\rho= \varepsilon \rho_1$
(for the case of a constant density star: $\rho(r)= e_0$). Each variable (except $\rho= e_0$ itself which is left
unexpanded) is now written as
\be
F= \varepsilon F_1 + \varepsilon^2 F_2 + O(\varepsilon^3)\, ; \quad {\rm etc.}
\ee

At second order, we {\it define}  the second-order values of the functions $V_{m0}$ and $V_{mk}$, by
(conventionally) using the same formulas as at first order, i.e.
\be
V_{m0(2)} \equiv -3V_2+2 \overline{Y}_2\;,\qquad V_{mk(2)} \equiv 2V_2- \overline{Y}_2\;.
\ee
We can use the inverse of these equations (see Eqs. \eqref{eqY1V1}) to express $V_2$ and $\overline{Y}_2$ 
in terms of $V_{m0(2)}$ and $ V_{mk(2)}$.

When expanding to second order our basic field equations $E_1$--$E_5$,\eqref{E1}--\eqref{E5}, we first get  algebraic
equations for $F_2$ and $\Lambda_2$ of the form
\bea
F_2 &=& V_{m0(2)}- 2\eta V_{mk(2)} + N_2^{F} \nonumber\\
\frac{1}{r}\Lambda_2 &=& V_{m0(2)} - \eta V_{mk(2)}  + N_2^{\Lambda} 
\eea
where   $N_2^{F}$ and $N_2^{\Lambda}$ are additional second-order contributions which are either quadratic in the
first-order variables $F_1, \Lambda_1, V_1, \bY_1$ (and, eventually, their derivatives), or linear in the pressure $P_2$.

We also get differential equations for $V_{m0(2)}$ and $ V_{mk(2)}$ of the form
\begin{eqnarray} \label{Vm2}
V_{m0(2)}^{\prime} &+& \frac{2}{r}V_{m0(2)}=S_{m0(2)}  \,,\nonumber\\
V_{mk(2)}^{\prime\prime} &+& \frac{2}{r}V_{mk(2)}^{\prime} - \left(\frac{2}{r^2}+\k^2\right)V_{mk(2)}=S_{mk(2)} \,,\nonumber\\
\end{eqnarray}
where the second-order source terms $S_{m0(2)}$ and $S_{mk(2)}$ consist of terms bilinear in $V_1$, $ \overline{Y}_1$, $F_1$, $\Lambda_1$, together with additional contributions linear in the pressure $P_2$ (remembering that $P$ is second-order, see section \ref{Linear}). We recall that $P_2$ is obtained by  solving the matter equation 
\be
P_2^{\prime}=-\rho F_1 \;,
\ee
with the condition that $P_2$ vanishes at the radius of the star $r=R_s$. 

The second-order solution is then explicitly obtained by using our general Green's function representation
\begin{eqnarray}
V_{m0(2)}(r)&=&\frac{1}{r^2}\int_{0}^{r}\hat{r}^2S_{m0(2)}(\hat{r})d\hat{r} \,,\\
V_{mk(2)}(r)&=&\int_{0}^{\infty}\hat{r}^2G_{\kappa}(r,\hat{r})S_{mk(2)}(\hat{r})d\hat{r} \,.
\end{eqnarray}
We found that it was possible to explicitly compute all the integrals generated by inserting the first-order
solution in the source terms $S_{m0(2)}$, $ S_{mk(2)}$ entering the latter second-order expressions.
The final expressions involve, besides elementary functions, some exponential-integral
functions $ {\rm Ei}(-x)$ with various arguments proportional to $z= \k r$ or $z_s= \k R_s$.
We recall that, with $x>0$,
\be
{\rm Ei}(-x) \equiv -\int_x^{\infty} dt \; \frac{e^{-t}}{t}\,.
\ee
It would take too much space to display here in full detail the second-order solutions (both in the interior
and in the exterior of the star) for all our variables.
We will only display here the function of most physical importance at the second order, namely the variable $F_2$,
which is the radial derivative of the second-order gravitational potential 
$\Phi= \varepsilon \Phi_1+ \varepsilon^2 \Phi_2 + O(\varepsilon^3)$.
As we shall explicitly discuss below, this is indeed the only variable whose second-order value is needed
to discuss the usual first post-Newtonian approximation. In addition, it is enough to know its value outside the star
to discuss its phenomenological implications as a modification of the usual Schwarzschild metric outside
a spherical mass distribution.

The full, second-order exterior solution $F_2$ has  a rather complicated structure, which can, however,
be explicitly displayed as follows:  
\begin{widetext}
\begin{eqnarray} \label{F2full}
F_2(r) &=& \frac{m_2(z_s)}{r^2}+\frac{2m_1^2}{r^3}  + \frac{e^{-z}(1+z)}{z^2}C_2^{F}(z_s)+ \frac{e^{-z}}{z^2}{\mathcal J}_0\left(z\right) + \frac{e^{-2z}}{z}{\mathcal P}_{0}\left(z\right) + \ln\left(\frac{z}{z_s}\right)\frac{e^{-z}(1+z)}{z^2}C_{LN}(z_s) \nonumber \\ 
& &  + \;{\rm Ei}(-z)\frac{e^{-z}(1+z)}{z^2}C_{E1}(z_s)  
 + {\rm Ei}(-2 z)\frac{e^{z}(z-1)}{z^2}C_{E2}(z_s)  + {\rm Ei}(-3 z)\frac{e^{z}(z-1)}{z^2}C_{E3}(z_s) \,, 
 \end{eqnarray}
where $z \equiv \k r$, $z_s \equiv \k R_s$, and where the dependence on the source characteristics of the various 
coefficients can be expressed in terms of two form factors:  
the previously defined form factor $\mathcal{F}(z_s)$, \eqref{Fz}, and a new one
denoted  $ \mathcal{E}(z_s)$ and defined as
\be \label{Ez}
 \mathcal{E}(z_s)=-\frac{e^{-2 z_s} }{z_s^5} \left(6  - 6  e^{2 z_s} + 12  z_s + 9  z_s^2 + 3  e^{2 z_s} z_s^2 + 3  z_s^3 -  e^{2 z_s} z_s^3\right) \,.
\ee
With this notation, the various terms in Eq. \eqref{F2full} are:
\begin{eqnarray} 
m_2(z_s)&=& \frac{\eta\,m_1^2 }{R_s}\mathcal{E}(z_s)  \label{Cm2} \\
C_{LN}(z_s)&=&-\frac{4}{3}\eta\, m_1^2 \kappa^3 \mathcal{F}(z_s) \label{CLN} \\
C_{E1}(z_s)&=& \frac{\eta(10-13 \, \eta)}{12}m_1^2\kappa^3\, \mathcal{F}^2(z_s) \\
C_{E2}(z_s)&=& -\frac{4}{3}\eta \,m_1^2\kappa^3\, \mathcal{F}(z_s) \\
C_{E3}(z_s)&=& C_{E1}(z_s) \\
{\mathcal P}_{0}\left(z\right) &=& \frac{m_1^2 \eta \kappa^3}{9  z^4} \left[-24(1+\eta) - 48 z(1+\eta) - 34 z^2(1+\frac{\eta}{2}) -  z^3(4-15 \, \eta)   \right. \nonumber \\
   & & \left.  + 
   16 z^4 \eta\right] \mathcal{F}^2(z_s) \\ 
 C_2^{F}(z_s) &=&  \frac{3 e^{- 
  2 z_s} m_1^2 \kappa^3 \eta}{16 z_s^6} (13 \,\eta-10) \left[-3 (1 + z_s)^2 + 
   \frac{4}{9} e^{2 z_s} z_s^6 \mathcal{F}^2(z_s)\right] {\rm Ei}(-z_s)   \nonumber \\
  & &   -\frac{4  \kappa^3 m_1^2 \, \eta}{3} \mathcal{F}(z_s) {\rm Ei}(-2z_s)  \nonumber \\
  & &   - \frac{3 m_1^2 \kappa^3 (z_s-1)}{16 z_s^6} (2 - e^{2 z_s} + 2 z_s + 
    e^{2 z_s} z_s) \eta (13 \,\eta-10) {\rm Ei}(-3z_s) \nonumber \\
  & &   -\frac{3 e^{ - 2 z_s}
   m_1^2 \kappa^3 (1 + z_s)^2 \eta ( 13 \,\eta-10)}{
 16 z_s^6} \left[{\rm Ei}(3z_s) -3  {\rm Ei}(z_s)\right] \nonumber \\
 & &   + \frac{e^{-3 z_s} \kappa^3 m_1^2 \eta }{12 z_s^7}\left[  -36 + 72 e^{2 z_s} - 36 e^{4 z_s} - 3 z_s + 102 e^{2 z_s} z_s - 
 99 e^{4 z_s} z_s + 39 z_s^2  \right. \nonumber \\
 & & + 294 e^{2 z_s} z_s^2 + 3 e^{4 z_s} z_s^2 - 
 6 z_s^3 + 240 e^{2 z_s} z_s^3 + 42 e^{4 z_s} z_s^3 - 12 z_s^4 - 
 40 e^{2 z_s} z_s^4 \nonumber \\
 & & - 12 e^{4 z_s} z_s^4 - 16 e^{2 z_s} z_s^5 - 
 108 \eta + 216 e^{2 z_s} \eta - 108 e^{4 z_s} \eta - 
 204 z_s \eta + 201 e^{2 z_s} z_s \eta \nonumber \\
 & & \left.+ 3 e^{4 z_s} z_s \eta - 
 72 z_s^2 \eta - 15 e^{2 z_s} z_s^2 \eta + 
 105 e^{4 z_s} z_s^2 \eta + 24 z_s^3 \eta + 
 60 e^{2 z_s} z_s^3 \eta - 36 e^{4 z_s} z_s^3 \eta  \right] \label{Q} \\
  {\mathcal J}_0(z)&=& 2\kappa^3 m_1^2 \eta \mathcal{F}(z_s) + \frac{2 \kappa^3 m_1^2 \eta}{3 z^3} \left[4 (1 + \eta) + 4 z (1 + \eta) + 
   z^2 (7 + \eta)\right] \mathcal{F}(z_s) \,.
  \end{eqnarray}
  \end{widetext}
  
 In order to better understand  the structure of $F_2$, let us study it under the two limits: (i) $r \to \infty$ at fixed $\k$ (so
 that $z = \k r \to \infty$); and (ii) $\k \to 0$ at fixed $ r> R_s$ (so that $z = \k r \to 0$ and $z_s = \k R_s \to 0$).
 The first limit studies the asymptotic structure of the solution at spatial infinity, while the second one would be the relevant
 one if (as is often done in massive-gravity studies) one would consider a Compton wavelength $\k^{-1}$ for the massive gravity
 excitation of cosmological size.

\subsection{Limit $r \to \infty$ at fixed $\k$ } 

Let us start by recalling that the first-order approximation to the exterior solution for $F= F_1+ F_2+ \cdots$ reads, according to \eqref{Flin},  as follows,
\begin{eqnarray}
F_1&=&\frac{m_1}{r^2} + \, \frac{e^{-z}(z+1)}{z^2}\;C_1^F(z_s) \,, \nonumber \\
C_1^F(z_s)&=&\frac{4 \k^2 m_1 \eta}{3}  \mathcal{F}(z_s) \,. \label{F1}
\end{eqnarray}
$F_1$ is the sum of a usual Newtonlike (and Schwarzschildlike) power-law contribution $m_1/r^2$, and of a decaying Yukawa contribution 
$\propto \d_r \left( e^{- \k r}/r\right)=- e^{- \k r}(1+ \k r)/r^2$.
Let us now consider the spatial asymptotics $r\rightarrow \infty$ of the second-order exterior solution $F_2$. 
To this end, we must take into account the asymptotic behavior of the exponential integral ${\rm Ei}(-z)$ 
(when $ z \to + \infty$) 
\be
{\rm Ei}(-z) \simeq -\frac{e^{-z}}{z}\left( 1-\frac{1!}{z}+\frac{2!}{z^2}+... \right) \;.
\ee
Using the latter asymptotic behavior, one concludes that $F_2$, \eqref{F2full}, contains four types of terms with  different behavior at infinity:
\begin{widetext}
\begin{eqnarray}
 \text{power-law:}&  &\frac{m_2(z_s)}{r^2}+\frac{2m_1^2}{r^3}  \\
  \propto e^{-z}\;:& &  \frac{e^{-z}(1+z)}{z^2}C_2^F(z_s)+ \frac{e^{-z}}{z^2}{\mathcal J}_0\left(z\right)  + {\rm Ei}(-2 z)\frac{e^{z}(z-1)}{z^2}C_{E2}(z_s) \\
   \propto e^{-z}\ln\left(\frac{z}{z_s}\right)\;:& & \ln\left(\frac{z}{z_s}\right)\frac{e^{-z}(1+z)}{z^2}C_{LN}(z_s) \\
   \propto e^{-2z}\;:& &  \frac{e^{-2z}}{z}{\mathcal P}_{0}\left(z\right) +
  {\rm Ei}(-z)\frac{e^{-z}(1+z)}{z^2}C_{E1}(z_s)  \nonumber \\
& &   + {\rm Ei}(-3 z)\frac{e^{z}(z-1)}{z^2}C_{E3}(z_s)  \;.
\end{eqnarray}
As a consequence the leading terms in the limit $r\rightarrow \infty$ of $F_1+F_2$ read,
\be
F_1+F_2=\frac{m_1+m_2}{r^2} + \frac{2m_1^2}{r^3}+\, \frac{e^{-z}(z+1)}{z^2}\;\left[ C_{1}^F(z_s)+C_2^{F}(z_s) + \ln\left(\frac{z}{z_s}\right)C_{LN}(z_s)  \right] +\, O\left(\frac{e^{-z}}{z^2}\right) +O\left( \frac{e^{-2z}}{z} \right) ,\label{F1+F2}
\ee
\end{widetext}
where $m_2 \equiv m_2(z_s)$ is given by Eq. \eqref{Cm2}, while  $C_1^{F}(z_s)$,  $C_2^{F}(z_s)$ and
$C_{LN}$  are given  by Eqs. \eqref{C1F}, \eqref{Q},\eqref{CLN}.

We see that if we define the total mass parameter $m$ of the star in torsion bigravity
(in the Schwarzschild sense of $ m = G M$, i.e. a length scale
associated with the mass) as the coefficient of $1/r^2$ in $F(r)$, as $ r\to \infty$
(i.e. $\Phi(r) \approx - m/r$ in this limit), we have
\be \label{mass}
m = m_1 + m_2 + O(\varepsilon^3)\,.
\ee
Here, we set  the bookkeeping parameter $\varepsilon$ back to 1 in the first two terms, but kept it in the error term as a reminder
that there are higher-order contributions that are at least cubic in the matter-density source $\rho$.

Before looking at the value of $m_2$ in various limits, let us note that the term $ \frac{2m_1^2}{r^3}$ 
is the second-order term in the $m/r$ expansion of a Schwarzschild solution, say $F_S(r)$, of mass $m_1$, indeed
\be
F_S(r)=\frac{m}{r(r-2m)} = \frac{m}{r^2} +  \frac{2m^2}{r^3}+ \cdots\;.
\ee
More generally, one can show by considering the  structure of perturbation theory in torsion bigravity
that, to all orders of perturbation theory, the asymptotic spatial behavior of the solution will be such that
the two independent (con-)torsion components \eqref{contorsion} are exponentially decaying (modulo
power-law and logarithmic factors),
\bea
 {K^{\hat{r}}}_{\hat{t}\hat{t}} &\underset{r \to \infty}=& O(e^{-\k r} )\,,\nonumber \\
 {K^{\hat{r}}}_{\hat{\theta}\hat{\theta}}= {K^{\hat{r}}}_{\hat{\phi} \hat{\phi}}&\underset{r \to \infty}=&  O(e^{-\k r})
 \eea
As a consequence, the variables $\Phi, F, \Lambda, V, W$ will asymptotically approach (modulo exponentially
small corrections) some Schwarzschildlike geometric
data (for some mass parameter $m$)
\bea
\Phi_S(r) &=&  +\frac12 \ln \left(1 - \frac{2 m}{r}\right) \nonumber\\
F_S(r) &=&\frac{m}{r(r-2m)} \nonumber\\
\Lambda_S(r) &=& -\frac12 \ln \left(1 - \frac{2 m}{r}\right) \nonumber\\
V_S(r)  &=& \exp[- \Lambda_S(r)] F_S(r) =\frac{m}{r^2}\left( 1-\frac{2m}{r} \right)^{-1/2} \nonumber\\
W_S(r) &=& - \frac{\exp[- \Lambda_S(r)]}{r}= -\frac{1}{r}\sqrt{1-\frac{2m}{r}} \,.
\eea

Let us look more closely at the value of the asymptotic mass $m = m_1 + m_2 + O(\varepsilon^3)$, 
and in particular at its second-order contribution $m_2(z_s)$. We recall that
\be
 m_1 =  G_0 M_{\rm bare}\,,
\ee
where $G_0 = 1/(16 \pi \lambda)$ is the (conventionally defined) massless spin-2 gravitational constant, 
and where
\be
M_{\rm bare} \equiv e_0 \times ({\rm volume}) = e_0  \frac{4 \pi R_s^3}{3} 
\ee
is the (conventionally defined) bare mass-energy of the constant-density star.
We  recall in this respect that in GR, the total Schwarzschild mass of a constant-density star is actually, simply
given by the Newtonlike expression
\be \label{mGR}
m_{\rm GR}=  G_N M_{\rm bare}=\frac{4 \pi G_N}{3} e_0 R_s^3
\ee
where $G_N$ denotes Newton's gravitational constant.
If we identify the torsion bigravity massless spin-2 gravitational constant $G_0 = 1/(16 \pi \lambda)$
with Newton's constant, $G_N$, we see that our first-order mass parameter $m_1$ (with units of length)
is equal to the (full) general relativistic mass parameter $m_{\rm GR}$.

On the other hand, the second-order contribution to the torsion-bigravity mass reads
\be \label{m2bis}
m_2 = \frac{\eta\,m_1^2 }{R_s}\mathcal{E}(z_s) 
\ee
where the form factor $\mathcal{E}(z_s)$ (with $z_s= \k R_s$) was defined in Eq. \eqref{Ez}.

We recall that the dimensionless parameter
\be
\eta = \frac{c_F}{c_R}= \frac{3}{4} \frac{G_m}{G_0}
\ee
is  a measure of the ratio between the coupling constant $G_m$ of the massive graviton and the coupling constant $G_0$
of the massless one. Therefore the ratio between $m_2$ and $m_1$ can be written as
\be \label{m2bym1}
\frac{m_2}{m_1}=  \frac{3}{4} \frac{G_m M_{\rm bare}}{R_s} \mathcal{E}(z_s) \,.
\ee
This expression is compatible with the idea that in the limit where $G_m/G_0 \to 0$ (at fixed $\k$) the torsion degrees
of freedom decouple from the matter so that torsion bigravity reduces to GR with $G_N=G_0$,
and the total mass parameter $m= m_1 + m_2 + \cdots$ reduces to its general relativistic value \eqref{mGR}.

It is interesting to discuss the physical consequences of the form factor $\mathcal{E}(z_s)= \mathcal{E}(\k R_s)$
entering $m_2$. It is easily checked that the form factor $\mathcal{E}(z_s)= \mathcal{E}(\k R_s)$ has the following
properties: (i) in spite of the prefactor $z_s^{-5}$ in its definition, $\mathcal{E}(z_s)$ is regular when $z_s \to 0$,
and has the finite limit
\be
\lim_{z_s\to0} \mathcal{E}(z_s)= - \frac{2}{5};
\ee
(ii) $\mathcal{E}(z_s)$ is negative in the interval $0 \leq z_s <  z_*$, and positive for $z_s > z_*$, where $z_* \approx 1.6969326$; 
and (iii) $\mathcal{E}(z_s)$ tends to zero like $+1/z_s^2$ when $z_s \to \infty$.

As a consequence of this behavior of the form factor $\mathcal{E}(z_s)$ we have the following limiting value
for $m_2$ as $\k \to0$ (i.e. $z_s \to 0$)
\be
m_2 \underset{\kappa \to 0}{\;\sim\;} -\frac{2m_1^2\eta}{5R_s}+\frac{2}{3}m_1^2\eta \kappa + ... \;.
\ee
[We will discuss the small $\kappa$ limit in more details in the next section.] The negative value of $m_2$
in this limit is probably due to the fact that the massive-gravitational binding energy $ - \frac{3}{5} G_m M_{\rm bare}^2/R_s$
(due to exchange of massive spin-2 excitations, in the small mass limit) dominates over other forms of binding energy
(e.g. pressure-related energy).

Another limit is the limit of very heavy massive spin-2 excitation ($\k \to \infty$), 
i.e. of a very short-range modification of gravity, $\kappa^{-1} \ll R_s $. In this case the second-order
correction to the mass parameter mass is found (as expected) to go to zero, 
\be
m_2 \underset{\kappa \to \infty}{\;\sim\;} \frac{m_1^2\eta}{R_s^3}\kappa^{-2} + O\left( \kappa^{-3} \right) \;.
\ee

\subsection{Limit $\kappa \rightarrow 0$ with fixed $r>R_s$}

Let us now study in more detail the limit where $\k$  becomes very small, i.e. where the Compton wavelength $1/\k$
is much larger than all the other scales of the problem (and notably $R_s$), being, e.g., of cosmological magnitude. 
This is the situation which is usually considered for massive gravity and bimetric gravity. As is well known
since the work of Vainshtein \cite{Vainshtein:1972sx}, the perturbation expansion of massive gravity (and
bimetric gravity) involves negative powers of  $\k^2$, which render the perturbative expansion invalid 
for radii $r$ smaller than some Vainshtein radius $R_V$ given, in generic (ghostfull) massive-gravity theories,
by the formula
\be \label{RV5}
R_V^{5} \sim \frac{G M}{\k^4} \sim \frac{m}{\k^4}\,.
\ee
More precisely, at the second-order approximation in $G$, the perturbative solution of the field equations of
generic massive-gravity (and bimetric gravity) theories contain terms 
that fractionally modify the linear approximation, say $\Phi_1 \sim m/r$ by terms of the type (see, e.g., \cite{Deffayet:2001uk})
\bea \label{Vexp}
\Phi &=& \Phi_1 + \Phi_2 + \cdots \sim \frac{m}{r} \left( 1 + \frac{R_V^5}{r^5} +\cdots \right) \nonumber\\
&\sim&  \frac{m}{r}  + \frac{m^2}{\k^4 \, r^6} +\cdots 
\eea
The latter expansion is performed in the domain $ R_s < r \ll \k^{-1}$, in the limit where $\k^{-1}$ is much larger $R_s$. 
[In this domain, and in this limit, one does not see the Yukawa exponential decay $\propto e^{- \k r}$.]

By contrast, we found the rather remarkable fact that, when considering the same limit, no terms involving inverse 
powers of $\k$ enter the perturbative expansion of torsion bigravity (in the domain $ R_s < r \ll \k^{-1}$)
up to the second order included.

For instance, the second-order contribution to $F$, considered in this limit, takes the following form
\begin{eqnarray} \label{F2outK0}
F_{2 \,\rm out}^{\kappa\rightarrow 0} &=& -\frac{2}{15}\eta(3+4\eta)\frac{m_1^2}{r^2R_s} + \frac{18+44\eta+25\eta^2}{9}\frac{m_1^2}{r^3} \nonumber \\
&-& \frac{4\eta(1+\eta)}{15}\frac{m_1^2R_s^2}{r^5}+O\left( \kappa \ln \kappa \right) \,,
\end{eqnarray}
where the $\k$-dependent piece tends to zero a $\kappa \to 0$. We have shown that, similarly,
 all the other field functions in second order perturbation theory, i.e. $V_2$, $ \overline{Y}_2$, $\Lambda_2$ 
 have finite limits (i.e. contain no denominators $\propto 1/\k^2$) as $\kappa \to 0$. 
 Such a result was {\it a priori} not all guaranteed because the field equations of torsion bigravity
 do contain  denominators $\propto 1/\k^2$. Indeed, such denominators come from the fact that the
 coefficient $c_{F^2}$ of the $F_{ij}^2$ terms in the action is proportional to $1/\k^2$, see Eq. \eqref{cRcFcF2}.

The absence of $O(1/\k^2)$ terms at second order is due to a special cancellation. Let us explain it.
We recall that the second-order variables $F_2$ and $\Lambda_2 $ are expressed in terms of the 
second-order potentials $V_{m0(2)} $ and $V_{mk(2)}$ via the equations
\bea
F_2 &=& V_{m0(2)}- 2\eta V_{mk(2)} + N_2^{F} \nonumber\\
\frac{1}{r}\Lambda_2 &=& V_{m0(2)} - \eta V_{mk(2)}  + N_2^{\Lambda} \,.
\eea
Here, the additional (nonlinear) terms $N_2^{F}, N_2^{\Lambda}$ 
(which are bilinear in $V_1$, $ \overline{Y}_1$, $F_1$, $\Lambda_1$ and their derivatives)
do contain some $1/\k^2$ factors, but all these
factors have a special structure: each monomial containing a factor  $\k^{-2}$ simultaneously contains at least one power
of $ \overline{Y}_1$ or of one of its derivatives. Similarly, the potentials $V_{m0(2)} $ and $V_{mk(2)}$
satisfy the differential equations \eqref{Vm2}
where the source functions $S_{m0\,(2)}$ and $S_{mk\,(2)}$ consist of terms bilinear in $V_1$, $ \overline{Y}_1$, $F_1$, $\Lambda_1$ and their derivatives. Again the latter bilinear expressions $S_{m0\,(2)}$, $S_{mk\,(2)}$ do contain some $1/\k^2$ factors, but the latter {\it a priori} dangerous (when $\k \to 0$) terms have the same special structure as $N_2^{F}, N_2^{\Lambda}$.  Each factor $\k^{-2}$
multiplies a monomial which is at least linear in $ \overline{Y}_1$ or one of its derivatives.

In turn, the reason why the terms $ \propto \k^{-2}  \overline{Y}_1 $, or  $ \propto \k^{-2}  \overline{Y}_1^{\prime} $, $\cdots$,
turn out to be innocuous in the limit $\k \to 0$
is that the variable   $ \overline{Y}_1$ happens to be of order $O\left(\kappa^2\right)$ as $\k \to 0$,
so that $ \k^{-2} \overline{Y}_1$ has a finite limit as $\k \to 0$. 
Indeed, from the definition \eqref{Vm0} one gets that
$$
 \overline{Y}_1=2V_{m0}+3V_{mk}  \;.
$$
Then, using Eq. \eqref{eqVmkGen} and the derivative of Eq. \eqref{eqVm0Gen}, one can see that $ \overline{Y}_1$ satisfies the following differential equation,
\be
\bY_1^{\prime\prime}+\frac{2}{r}\bY_1^{\prime}-\frac{2}{r^2}\bY_1=3\kappa^2 V_{mk}+2S_{m0}^{\prime}+3S_{mk} \;.
\ee
At the linear level, the source terms $S_{m0(1)}$, $ S_{mk(1)}$, read, according to Eqs. \eqref{eqVm0}, \eqref{eqVmk},
\begin{eqnarray}
 S_{m0(1)}&=&\frac{ \rho(r)}{4\lam} \;, \nonumber\\
 S_{mk(1)} &=& - \frac{\rho^{\prime}(r)}{6\lam} \,,
 \end{eqnarray}
 so that the combination of source terms entering the equation for $\bY_1$ cancells:
 \be
 2S_{m0(1)}^{\prime}+3S_{mk(1)}=0 \label{S+S=0}\,.
\ee
Finally, $\bY_1$ satisfies an equation whose right-hand-side is explicitly $O(\k^2)$, namely
\be
\bY_1^{\prime\prime}+\frac{2}{r}\bY_1^{\prime}-\frac{2}{r^2}\bY_1=3\kappa^2 V_{mk}\;.
\ee
This explains why $\bY_1$ is of order $O(\kappa^2)$, thereby ensuring the absence of denominators $1/\k^2$
in the second-order solution.

It is not {\it a priori} clear whether this (or a similar) cancellation mechanism will continue to work at the third order
of perturbation theory. [The specific property \eqref{S+S=0} does not seem to persist for $S_{m0(2)}$ and $S_{mk(2)}$.]
We note that one cannot apply the same reasonings to the next (third) order of perturbations because the property \eqref{S+S=0} is not true for $S_{m0\,(2)}$ and $S_{mk\,(2)}$. This means that it is a priori possible that the perturbation theory
will involve $1/\k^2$ factors in the third order. We leave the investigation of this subject to future work, 
and comment below on what would be the
consequences of  the presence of $1/\k^2$ factors at the third order of perturbation theory. For the time being, we shall
continue studying the consequences of our results at the second order of perturbation theory.

\section{Numerically constructing exact star solutions}

In GR, it is possible to write down analytically the exact solution for
the metric generated by a constant-density perfect fluid \cite{Schwarzschild:1916ae}.
Though the exterior Schwarzschild solution \cite{Schwarzschild:1916uq} is an exact
exterior solution of torsion bigravity (with zero contorsion), this is {\it not true} for
the interior Schwarzschild solution. Indeed, as we saw in our perturbation theory
analysis, the presence of a nonzero $T_{\mu \nu}$ in space, necessarily generates
some nonzero contorsion field, i.e. a difference between the affine connection $A^i_{j \mu}$
and the Levi-Civita connection ${\omega^i}_{j\mu}$. And indeed, one can check
that the interior Schwarzschild solution (with zero contorsion) does not satisfy the
field equations of torsion bigravity.

As the analytic construction of an exact analytical solution of the complicated system 
of torsion bigravity spherically-symmetric field equations discussed in Sec. \ref{NonlinSys}
seems difficult, we have appealed to numerical methods to confirm the global existence
of regular solutions of torsion bigravity satisfying the boundary conditions imposed in our
perturbation theory. Let us recall that these boundary conditions are: (1) geometric regularity
of all our variables at the origin $ r \to 0$, and (2) decay of all our variables at spatial infinity $r \to \infty$.

We recall that the system of equations to be satisfied (in presence of matter) consists either of: (i) the original six
field equations comprising $E_1$--$E_5$, together with the matter equation $E_m$ (knowing that this system is constrained
by two other equations that must be satisfied); or (ii) a reduced system made of the three radial-evolution equations
\eqref{sys3}, plus the radial-evolution equation \eqref{Peq} for the pressure $P(r)$.
In our numerical simulations, we have used the reduced first-order system of four ordinary differential equations (ODE's)
defined by Eqs. \eqref{sys3}, \eqref{Peq}, for the four variables $V, \bY, \pi, P$. This system is completed
by giving an equation of state for the matter. In our simulations we use the simple condition of constant density:
$\rho(r)=e_0$. After integrating this system, the
values of the variables $F$, $L$ (or $\Lambda$), and $W$ was obtained by using
Eqs. \eqref{Fsol}, \eqref{Lsol}, and \eqref{defs}.

As we have seen in Sec. \ref{Linear}, in perturbation theory the boundary conditions (1) and (2) (together with
the choice of the radius $R_s$ of the star) uniquely determine (at each order
of perturbation theory)  a torsion bigravity solution. The main motivation
for constructing numerical solutions was to prove that this uniqueness property holds in the full nonlinear theory.
To do this we need to study what the conditions of regularity at the origin impose as constraints 
on the initial conditions (at $ r \to 0$) of our four variables $V(r), \bY(r), \pi(r), P(r)$.
First, the geometric character (scalar, vector, tensor, etc) of our variables show that,
near the origin, they must admit general Taylor expansions of the following restricted type:

\begin{eqnarray}
V(r) &=& v_1r+v_3r^3+O\left( r^5 \right),  \nonumber\\ 
\overline{Y}(r) &=& y_1r+y_3r^3+O\left( r^5 \right),  \nonumber\\ 
\pi(r) &=& \pi_0+\pi_2r^2+O\left( r^4 \right),  \nonumber\\ 
P(r) &=&  P_0+P_2r^2+O\left( r^4 \right)\,, \label{InitialCond}
\end{eqnarray}
together with
\begin{eqnarray} \label{InitialCondFL}
F(r) &=& f_1r+f_3r^3+O\left( r^5 \right),  \nonumber\\ 
\Lambda(r) &=& \Lambda_2r^2+\Lambda_4r^4+ O\left( r^6 \right)\,. 
\end{eqnarray}
[$\Lambda( 0) =0$ is necessary to have a locally flat metric at the origin.]
By inserting these expansions into the equations of our system, we get, at each order in $r$
some relations between the various expansion coefficients. The crucial point is that, if we consider the central value $P_0=P(r=0)$
of the pressure as a given quantity (that will determine the radius, given the constant density $e_0$), the equations of our
system give enough relations to determine all the other expansion coefficients $v_n, y_n, \pi_n, P_n$ in terms of 
{\it only one of them}.
We have chosen $v_1$ {\it as unique free inital datum}. For instance, at the lowest order in the $r$ expansion, one finds
that $y_1$, $\pi_0$,  $f_1$ and $\Lambda_2$ are determined by $v_1$ and $P_0$ (and $e_0$) by the following
formulas
\bea
 y_1 &=& \frac{1}{24 \lam} (e_0 - 3 P_0 + 36 \lam v_1) \nonumber \\
\pi_0 &=& \frac{1}{12\lam} (e_0 - 3 P_0 + 24\lam v_1) \nonumber \\
 f_1 &=& \frac{1}{12(\lam-c)}(e_0+3P_0-12c  v_1)   \nonumber \\
 \Lambda_2 &=& \frac{1}{24\lam(c-\lam)}\left[(c-2\lam) e_0 - 3c P_0+ 12 c \lam v_1\right] \nonumber\\ 
\eea
Similar formulas also determine the next order coefficients in the $r$ expansion:  $v_3$, $y_3$, $\pi_2$, $P_2$, etc.

In other words, a single ``shooting parameter" at the origin, namely $v_1$, uniquely determines (after having
chosen $P_0$) the solution of torsion bigravity. When integrating the system, the value $R_s(P_0, v_1)$
of the star radius will be obtained as the (first) radius where $P(r)$ (vanishes). 

For $r > R_s(P_0, v_1)$ one sets $\rho(r)=0$ and $P(r)=0$ and
continues  integrating the three field equations \eqref{sys3} to get the exterior solution
for the three variables $V, \bY, \pi$.
For a generic value of $v_1$, the so-constructed exterior solution for $V, \bY, \pi$ (and the associated
values of $F$, $W$ and $\Lambda$) will {\it not} decay at infinity, but will contain some growing exponential pieces
$\propto e^{+ \k r}$. We have seen in Sec. \ref{Linear} that the general exterior solution contains
three parameters: one parameter, say $m$, (Schwarschild-type total mass) parametrizing all the
power-law behaviour of the solution (asymptotically described by a Schwarzschild metric and connection);
together with two parameters, say $C_+$ and $C_-$ respectively parametrizing the exponentially growing,
and decaying, Yukawa-type contributions to the solution. At the linear level, each variable contains
different coefficients $C_+$ and $C_-$, e.g.,
\be
F_1(r) \approx \frac{m}{r^2}+C_{-}^F \frac{e^{- \k r}(1+ \k r)}{r^2} +  C_{+}^F \frac{e^{+ \k r}(1- \k r)}{r^2}\,,\label{Fexp}
\ee
but all the exponential-mode coefficients are related between themselves by the field equations, so that
only two of them are independent. 

In order to satisfy the decaying boundary condition at spatial infinity, we finally have a one-parameter shooting
problem, namely it is enough to impose that (given some value of $P_0$) the coefficient $C_+(v_1)$  of one variable vanishes.
To numerically extract from numerical data an estimate of the (common, underlying) $C_+(v_1)$ coefficient, we worked with
the variable $V_{mk}(r)\equiv 2V(r)- \overline{Y}(r)$ which does not contain a mass-type, power-law contribution.
In practical terms, this meant tuning the value of $v_1$ at $r=0$ until reducing essentially to zero
the value of, say,
\be
C_+^{\rm eff}(r_0) \equiv \frac{V_{mk}(r_0)}{e^{+ \k r_0}(1- \k r_0) r_0^{-2}}\,,
\ee
taken at some large value of $r_0$ (such that $ e^{+ \k r_0} \gg1$, so that the exponentially decaying
contribution to $V_{mk}(r_0)$ is fractionally negligible). [In practice, we used $\k r_0=10$
corresponding to $ e^{+ \k r_0} \approx 2 \times 10^4$]. The tuning of $v_1$ is obtained by a
simple dichotomy procedure, i.e. alternating the signs of $C_+^{\rm eff}(r_0; v_1)$ by changing the
value of $v_1$ until $C_+^{\rm eff}(r_0; v_1)$ is smaller than what
is permitted by the numerical accuracy of our simulation.

We implemented this simple, one-parameter shooting strategy for several star models, of various radii
and compactnesses. Let us only indicate here our results for one such star model.
Without loss of generality, we used units where $\kappa=1$ and $\lam=1$. The first condition says that we 
measure lengths in  units of $\kappa^{-1}$, while the second one defines the (independent) unit for
the Newtonian constant such that $16 \pi G_0=1$. Here, we shall exhibit a specific star model having the following
physical characteristics. First we set the dimensionless torsion bigravity parameter $\eta$ to the
value $\eta=1$, i.e. $c_F=c_R$ (both being equal to $\frac12$ in our units where $\lam=c_F+c_R=1$).
The other physical choices concern: (a) the radius of the star in units of $\kappa^{-1}$, i.e. the dimensionless
quantity $z_s = \k R_s$, and (b) the value of the star compactness\footnote{We normalize the definition of the compactness so that it is equal to 1 for a black hole in GR. See below the exact definition of $ C_s$.},  $ C_s  \simeq 2 G_0 M_s/R_s$, with
$M_s \equiv \frac{4 \pi}{3} e_0 R_s^3$. The two quantities $z_s$ and $C_s$ are dimensionless, and physically
depend on the two independent values of $e_0$ and $P_0$. We have chosen (in our units) the specific values
\bea \label{numdata1}
e_0 &=& 3. \nonumber\\
P_0 &=& 0.866020112678
\eea
These values were chosen by using, as guideline, our perturbation-theory expressions, with the aim of getting
a star model having $\k R_s \sim 1$, and a sufficiently high compactness $C_s \sim 0.3$ (comparable
to the expected compactness of a  neutron star in GR).

Anyway, after doing the choices \eqref{numdata1}, we found that we needed to tune $v_1$ to the value
\be \label{numdata2}
v_1^{\rm tuned} \approx 0.05367018\,,
\ee
to get a sufficiently small value of $C_+^{\rm eff}(r_0)$, i.e. a solution exhibiting numerical decay
up to $ r\sim 10/\k$. As said above, we obtained $v_1^{\rm tuned}$ by dichotomy, using
as first guesses the analytical estimates of $v_1$ obtained either directly from linearized perturbation theory, namely
\be
v_1^{\rm lin}= \frac{m_1}{R_s^3}\left[ 1 - \frac43  \, e^{-\kappa R_s}(1+\kappa R_s)  \right]\,,
\ee
or, alternatively, by combining the relation between $v_1$ and the
value $f_1$ of $F^{\prime}\vert_ {r\to0}$ with the analytical estimate for $f_1$  deduced
from our linear linear solution \eqref{Flin}, i.e. 
\be
f_1^{\rm lin}= \frac{m_1}{R_s^3}\left[ 1 + \frac43 \eta \, e^{-\kappa R_s}(1+\kappa R_s)  \right].
\ee 
The numerical solution was found to have a star radius equal (in our units
where $\k=1$)
\be
R_s \approx 0.739525\,.
\ee
The value of the star radius was numerically determined by looking at the point where the pressure $P(r)$ vanishes.

\begin{figure}
\includegraphics[scale=0.5]{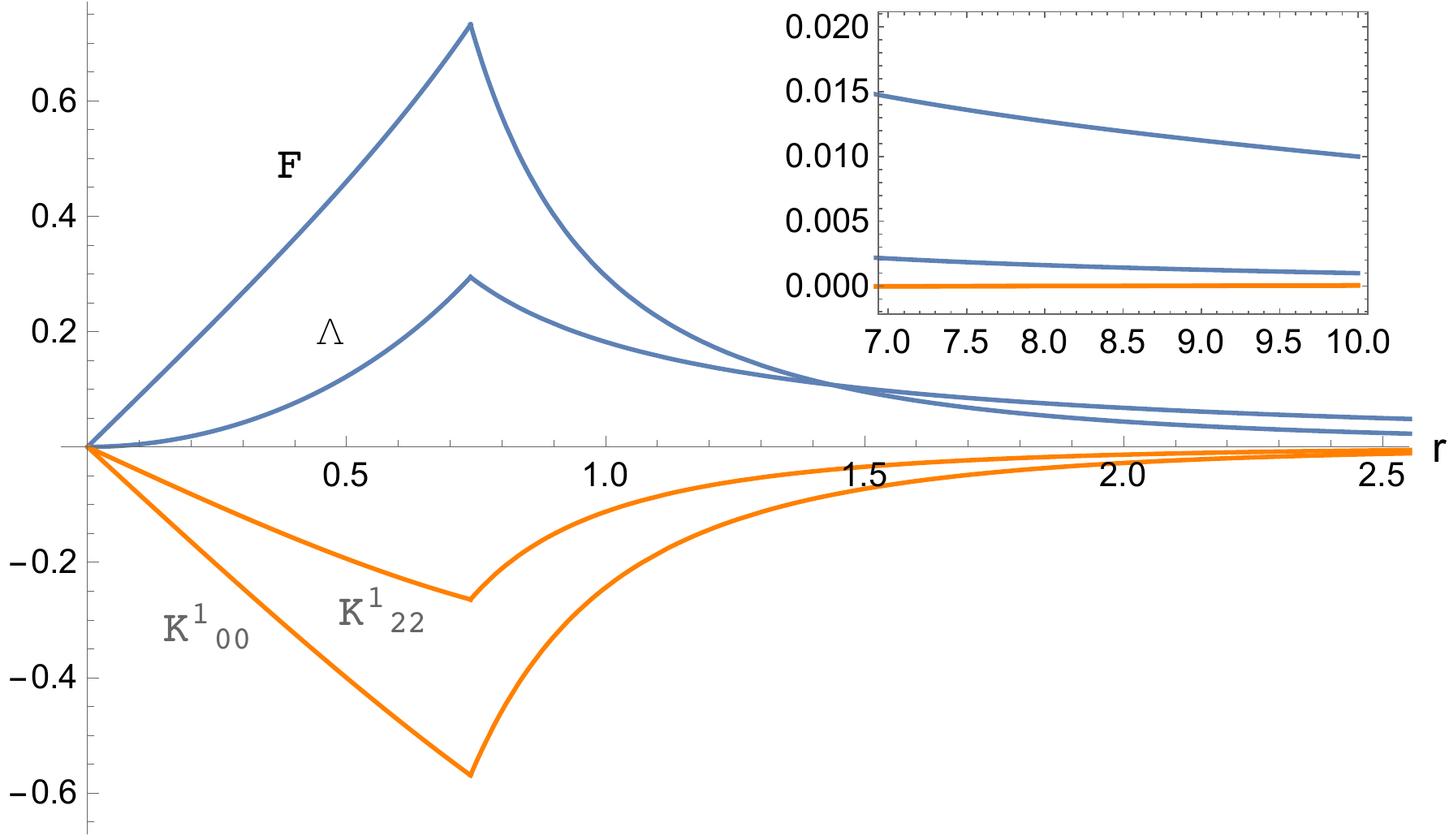}
\caption{\label{fig3}
Starting from the top left, one displays four functions characterizing our numerical star solution: 
the two independent metric functions $F(r)\equiv \Phi^{\prime}(r)$, $\Lambda(r)$,  and  
(in the lower part of the graph) the two independent (con-)torsion components 
$ {K^{\hat{1}}}_{\hat{0}\hat{0}}$  and $ {K^{\hat{1}}}_{\hat{2}\hat{2}}$. The inset contrasts the power-law
decay of the metric functions with the exponential decay of the torsion ones.
}
\end{figure}

We display in Fig. \ref{fig3} the numerical values (both inside and outside the star)
of four variables encapsulating the essential
geometrical properties of our solution, namely: $F(r)$, $\Lambda(r)$, and the two independent (con)torsion components,
namely $ {K^{\hat{1}}}_{\hat{0}\hat{0}}$  and $ {K^{\hat{1}}}_{\hat{2}\hat{2}}$,
as defined in Eq. \eqref{contorsion}. While $F(r)$ and $\Lambda(r)$ decay for large $r$ in a power-law fashion
($F(r) \propto 1/r^2$ and $\Lambda(r) \propto 1/r$), the torsion components decay exponentially. 
Note that the order of magnitude of the torsion inside the star is comparable to the value of $F$.
As ${K^{\hat{1}}}_{\hat{0}\hat{0}} = V- e^{-\Lambda} F$ (from \eqref{contorsion}), we see
that the matter density of the star generates a torsion which is of roughly the same magnitude
as the component ${\omega^{\hat 1}}_{\hat {0} \hat{0}} = e^{-\Lambda} F$ of the Levi-Civita
connection. [From Eqs. \eqref{Klin}, this remains true even when $\eta \to 0$.]
 
In order to measure the deviation from GR implied by our numerical star model, we have extracted
several observable, gauge-invariant characteristics of our solution.
First, we extracted an estimate of the total Keplerian-Schwarzschildian mass parameter $m_S$ (as measured faraway) 
by fitting (in the interval $6< r < 10$) the numerical value of $r^2 F(r)$ to its analytically predicted asymptotic
expansion $\sim m_S (1+2 m_S/r) + C_{-}^F e^{- \k r}(1+ \k r) +  C_{+}^F e^{+ \k r}(1- \k r)$.
This gave us
\be \label{mS}
m_S=0.1005(3) \;.
\ee
where the digit in parenthesis is a rough measure of the uncertainty (in the last digit) on the numerical determination of $m_S$.
Note that this is only slightly smaller than what would be the value of the total mass in Einstein's theory,
namely $m_{\rm GR} = e_0 R_s^3/12 \approx 0.101111$. We have verified that such a value is compatible with
our second-order-corrected mass value, $m_1+m_2$, with $m_2$ given by Eq. \eqref{m2bym1}. [It happens that the
form factor $\mathcal{E}(z_s)$, though still negative, is quite small, thereby explaining why one does not see the
expected larger self-gravity binding effect due to a high compactness $\sim 0.3$.]



The formally defined compactness $2 m_S/R_s$ would then be $2 m_S/R_s\approx 0.272$.
However, such a formal definition (directly copied on GR expressions) does not
correspond to any observable characteristics of a star in torsion bigravity.
We therefore extracted other (in principle) observable features and numbers from our solution.

\begin{figure}
\includegraphics[scale=0.7]{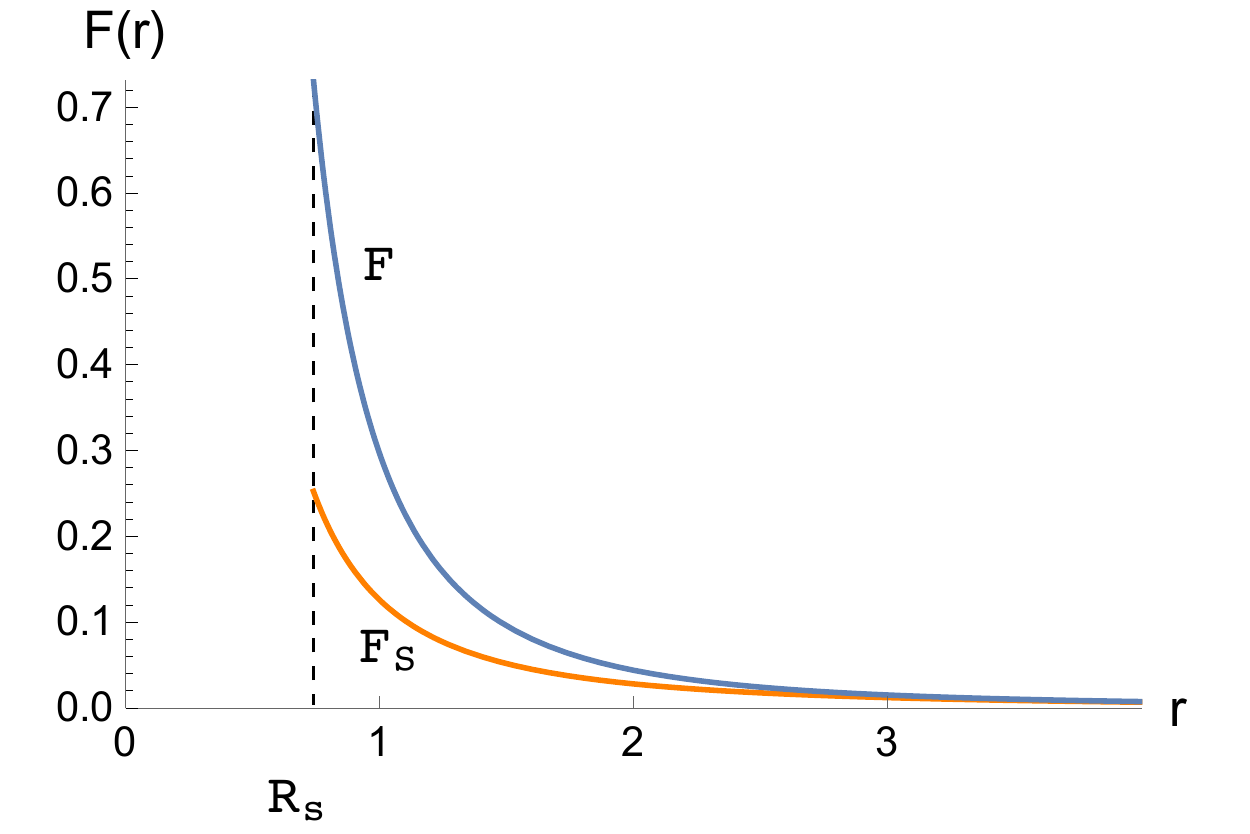}
\caption{\label{fig6}
Comparing $F(r)$ (upper curve; blue online), for our exterior solution, with  $F_S(r)$, corresponding to an exterior 
Schwarzschild solution with the 
same asymptotic Keplerian mass $m_S$, Eq. \eqref{mS} (lower curve; orange online). }
\end{figure}

\begin{figure}
\includegraphics[scale=0.7]{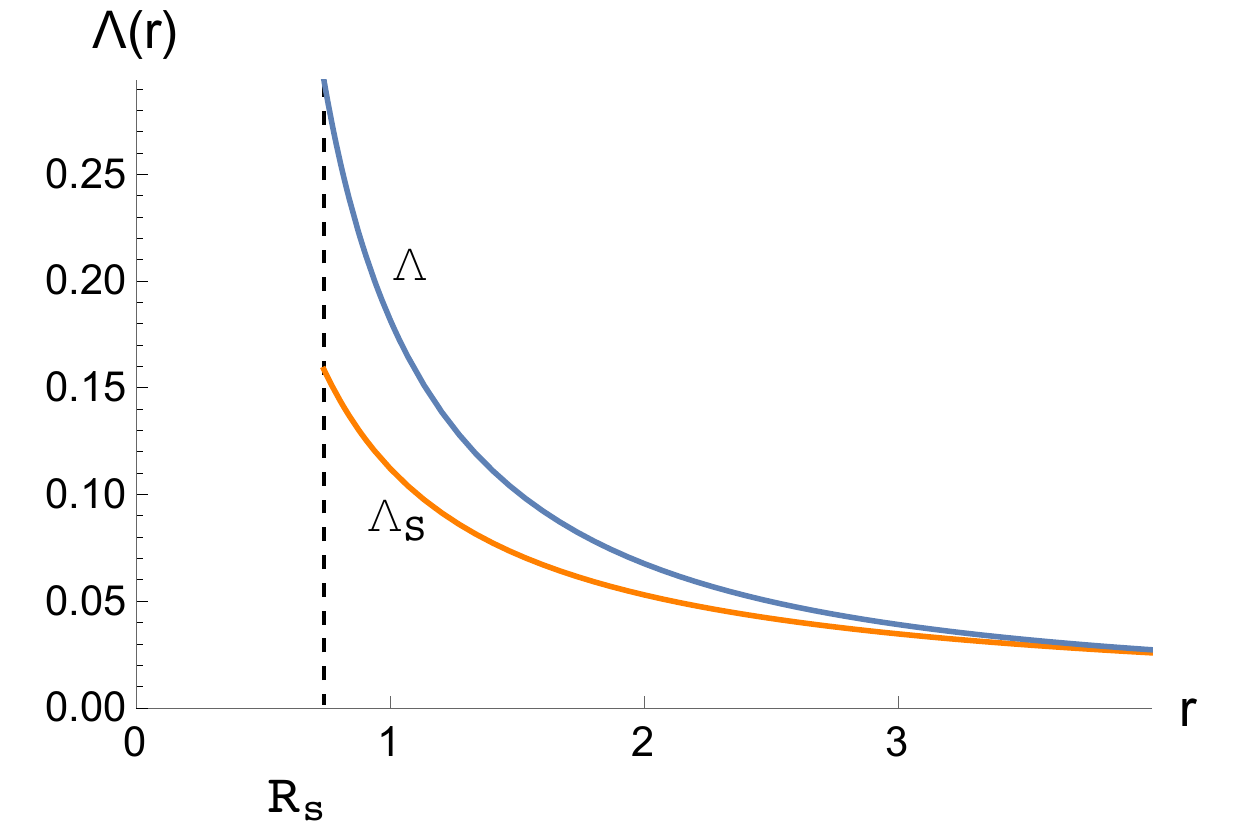}
\caption{\label{fig7}
Comparing $\Lambda(r)$ (upper curve; blue online), for our exterior solution $r \geq R_s$, with $\Lambda_S(r)$, corresponding to 
an exterior Schwarzschild solution 
with the same asymptotic Keplerian mass $m_S$, Eq. \eqref{mS} (lower curve; orange online). }
\end{figure}

We have seen above that if one probes our bigravity field at, say, distances $r \gtrsim 5/\k$, the geometry
will look like a GR metric of mass $m_S$. On the other hand, the exact torsion bigravity metric functions 
$F= \Phi^{\prime}$ and $\Lambda$ start significantly differing from their GR counterparts
$F_S(r)\equiv F_{\rm GR}(r, m_S)$ and $\Lambda_S(r)\equiv\Lambda_{\rm GR}(r, m_S)$
as $r$ gets smaller and comparable to $1/\k$. This is illustrated in Figs. \ref{fig6} and \ref{fig7}.
These figures show that, near the star, the torsion bigravity solution differs by $ \gtrsim 100\%$ 
from its GR counterparts.


Let us observationally define the compactness of a star by the surface value of
\be
{\cal C}_s \equiv 1- e^{2 \Phi(R_s)} \; ( = 2 GM/R_s \; {\rm in \; GR})
\ee
In our torsion bigravity model, we can compute the surface value $\Phi(R_s)$ of $\Phi$ 
(relative to a zero value at infinity: $\Phi(\infty)=0$) by integrating $F(r)$, namely
\be
\Phi(R_s)  = -\int_{R_s}^{\infty}{Fdr} \;.
\ee
A numerical evaluation of this integral gave us
\be
\Phi(R_s)\approx - 0.302028 \;,
\ee
and therefore
\be \label{CsTBG}
{\cal C}_s \equiv 1- e^{2 \Phi(R_s)} \approx 0.453410
\ee
Note that this is significantly larger than the corresponding value in GR for a star having the same mass
and the same radius, namely
\be \label{CsGR}
{\cal C}_s^{\rm GR} = \frac{2 m_S}{R_s}= 0.271796
\ee
As a supplementary measure of the strong-gravity nature of our torsion-bigravity star model, let us also
cite the value of the geometric-deformation quantity $1- e^{-2 \Lambda(R_s)}$ (which is also equal to 
the compactness $\frac{2 m_S}{R_s}$ in GR)
\be
1- e^{-2 \Lambda(R_s)} = 0.444855 \,. 
\ee
This value confirms that our torsion bigravity model induces large deformations of the geometry.

Another quantity of direct observational significance is the radius of the  innermost (or last) stable circular orbit (LSO). From
 Eq. (4.14) of Ref. \cite{Damour:2009sm}, the condition defining the LSO reads (in terms of the
 variables $A \equiv e^{2\Phi}$ and $ u\equiv \frac{1}{r}$) ,
\be
2A \frac{\d A}{\d u} + 4 u \left( \frac{\d A}{\d u} \right)^2 - 2 u A \frac{\d^2 A}{\d u^2} =0\;.
\ee
Transcribed in terms of the function $F(r)$, this yields
\be
-3F(R_{\rm LSO})+2R_{\rm LSO}F^2(R_{\rm LSO}) - R_{\rm LSO}F^{\prime}(R_{\rm LSO})=0 \,.
\ee
Solving this equation gave us 
\be
R_{\rm LSO} \approx 1.549 \approx 15.4 m_S   
\ee
Note that the ratio $R_{\rm LSO}/m_S \approx 15.4$ is about 2.57 {\it larger} than the well-known corresponding
GR value $R^{\rm GR}_{\rm LSO}/m_S =6$. This difference is linked to the fact
(already apparent in Fig.  \ref{fig6}) that the gravitational field near a torsion bigravity star
(of a given Keplerian mass) is significantly more attractive than in GR. [This increase in the strength
of the gravitational attraction is essentially due to the extra (short-range) attraction provided by
the massive spin-2 excitation.]  Note that the value of the ratio $R_{\rm LSO}/m_S$ is in principle
extractable from the observation of an accretion disk around a neutron star. 

In the following section, we shall discuss more potential phenomenological aspects of torsion bigravity.

\section{Phenomenology of torsion bigravity} \label{phenom}

We present a preliminary analysis of the phenomenology of torsion bigravity based on the first two orders of perturbation theory,
and focussing on solar-system tests of gravity.
\subsection{Assuming $ r \sim 1/\k$}

Let us first consider the case where $ r \sim 1/\k$, i.e. when the exponential decrease of the massive
spin-2 excitation is important in the considered physical situation. In that case, torsion bigravity
already introduces a modification of Einstein's (purely massless) theory at the Newtonian level, i.e.
when considering the linearized-gravity interaction between two slowly moving massive objects.
As already mentioned, previous studies of the linearized approximation \cite{Hayashi:1980ir,Nikiforova:2009qr}
have shown that the linearized interaction between two massive objects (with stress-energy tensor $T_{\mu \nu}$)
involves the exchange of two fields: a massless Einsteinlike gravitational field $h^*_{\mu \nu}$, and a
massive spin-2 field (contained within the 24 components of the contorsion tensor). 
The massless field $h^*_{\mu \nu}$ couples to $T_{\mu \nu}$ with the Newtonianlike
coupling constant
\be
G_0 = \frac{1}{16 \pi \lambda}= \frac{1}{16 \pi (c_R + c_F)}\,,
\ee
while the massive spin-2 excitation couples to $T_{\mu \nu}$ with the effective Yukawa-Newtonian
coupling constant
\be
G_m = \frac43 \eta \, G_0= \frac43 \frac{c_F}{c_R} \, G_0
\ee
This means that the gravitational interaction term of the source $T_{\mu \nu}$ with itself (after
integrating out the field degrees of freedom) reads
\be
S_{\rm int} = \int d^4 x L_{\rm int}
\ee
with
\bea
 L_{\rm int}&= &   2\, G_0 T^{\mu \nu} \left( \frac{- 4 \pi}{\Box} \right) \left( T_{\mu \nu} - \frac12 T \eta_{\mu \nu} \right) \nonumber \\
&+& \frac32 G_m T^{\mu \nu} \left( \frac{- 4 \pi}{\Box - \k^2} \right) \left( T_{\mu \nu} - \frac13 T \eta_{\mu \nu} \right). \nonumber \\
\eea 
Here the extra numerical prefactors $2$ and $\frac32$ are such that the interaction between two nonrelativistic
($T_{\mu \nu}= T_{00} \delta^0_\mu \delta^0_\nu$) stationary ($\Box= \Delta$) sources read
\bea
L_{\rm int}^{\rm Newtonian} &=&    G_0 T_{00} \left( \frac{- 4 \pi}{\Delta} \right) T_{00} \nonumber \\
&+&  G_m T_{00} \left( \frac{- 4 \pi}{\Delta - \k^2} \right) T_{00}
\eea 
If we consider the interaction between a test particle of mass $M_2$ and a spherical object 
(say a nonrelativistic star) of constant density $e_0$ and total mass $M_1= \int d^3x e_0$,
separated by a distance $r_{12}$ (between their centers of mass), the above formulas yield an interaction
potential $V_{\rm int} = -\int d^3x L_{\rm int}$
\be \label{Vint}
V_{\rm int}^{\rm Newtonian}= -G_0 \frac{M_1 M_2}{r_{12}}- G_m \mathcal{F}(\k R_1) \frac{M_1 M_2 e^{- \k r_{12}}}{r_{12}}.
\ee
Here the form factor $\mathcal{F}(\k R_1) $ (where $R_1$ denotes the radius of the object $M_1$) is the (normalized) one
introduced in Eq. \eqref{Fz}. [If we were considering the interaction between two constant-density spherical objects,
we should include two form factors: $\mathcal{F}(\k R_1) \mathcal{F}(\k R_2)$. In the case of a test particle
considered here, we have $\mathcal{F}(\k R_2) \to 1$.]
It is easily checked that the radial force $ F_{\rm int} =- \d V_{\rm int}^{\rm Newtonian}/ \d r_{12}$
deduced from the interaction potential is simply equal to (setting $z_s = \k R_1$)
\bea
 F_{\rm int}&=& -M_2 \left( G_0 \frac{M_1}{ r_{12}^2} + C_1^F(z_s) \, \frac{e^{-\k r_{12}}(1+ \k r_{12})}{(\k r_{12})^2} \right) \nonumber\\
 &=& -M_2 F_1(r_{12})\,,
\eea
where the function $F_1(r)$ denotes the external value of our linearized variable $F(r) = \Phi^{\prime}(r)$,
as obtained in Eq. \eqref{Flin} above. This is a direct check of the superposition of massless and massive spin-2
excitations in the Newtonianlike potential $\Phi = \frac12 \ln ( -g_{00})$.

There are many experimental data that have set upper limits on the existence, in addition to the Newtonian
$1/r$ interaction, of a Yukawa-type interaction $\alpha \, e^{-\k r}/r$ coupled with gravitational strength
to matter. See Refs  \cite{Fischbach:1999bc,Adelberger:2003zx} for reviews of the experimental situation.
[Note that, when considering non spin-polarized sources, the torsion bigravity interaction respects
the equivalence principle, as assumed in the presently considered {\it composition-independent} limits.]
The Yukawa strength parameter $\alpha$ entering these limits is simply $\alpha= G_m/G_0= \frac43 \eta$.
The experimental limits on $\alpha$, as a function of $\lambda \equiv 1/\k$ are summarized in Fig. 
2.13 of \cite{Fischbach:1999bc} and Fig. 4 of \cite{Adelberger:2003zx}
(for the range $10^{-3} {\rm m} < \lambda < 10^{+ 15} {\rm m} $).
We note that the less stringent upper limits apply in the geophysical range (i.e. for 
$ 1 \, {\rm m} \lesssim \k^{-1} \lesssim 10 \, {\rm km}$) and roughly limits $\eta= \frac34 \alpha$,
to be
\be
\eta \lesssim 3 \times 10^{-4} \; \;  {\rm for} \; \; \k^{-1} \lesssim 10 \, {\rm km}\,.
\ee
A range of order $ \k^{-1} \sim 10 \, {\rm km}$ is interesting to consider if one wishes
to discuss possible deviations from GR in the physics of neutron stars and black holes.

\subsection{PPN parametrization of the second-order torsion bigravity metric when assuming $R_s < r \ll 1/\k$}

Let us now consider the other phenomenological situation where the massive-gravity range is much larger
than all the length scales of our system. [We exclude from our consideration the case where
$1/\k$ is roughly between $10 \, {\rm km}$ and $10^{+ 11} {\rm km} $, for which there are very stringent
limits on $\eta$ coming from Earth-satellite, lunar and planetary data.]

If we consider the motion of classical, non-spin-polarized, test masses in our second-order torsion bigravity
spacetime (endowed with the metric $g_{\mu \nu}$ and the connection $A_{i j \mu}$), it is given
(as shown in Ref. \cite{Hayashi:1980av}) by geodesics of the metric $g_{\mu \nu}$. The observational
differences (say for the motion of the planets around the Sun) between torsion bigravity and  GR are then
encapsulated in the difference between our spherically symmetric metric
\be \label{ds2bis}
ds^2=-e^{2\Phi}dt^2 + e^{2\Lambda}dr^2 + r^2\left( d\theta^2+\sin^2\theta\, d\phi^2 \right) \; \;,
\ee
and the usual Schwarzschild metric. As is well-known, solar-system experiments are primarily sensitive only
to the first post-Newtonian approximation to the metric in the solar system, which is described by the
Eddington PPN parameters $\beta$ and $\gamma$. When using (as we do) a Schwarzschildlike radial
coordinate, the PPN parameters are defined by writing the first post-Newtonian metric as 
(see, e.g., \cite{Weinberg:1972kfs})
\begin{eqnarray} \label{PPNmetric}
-g_{00}^{\rm PPN} & =& e^{2\Phi} = 1-\frac{2m_0}{r} + 2(\beta-\gamma)\frac{m_0^2}{r^2} + O\left[ \frac{m_0^3}{r^3}\right] \,,\nonumber \\
g_{rr}^{\rm PPN} &=& e^{2\Lambda} = 1+2\gamma \frac{m_0}{r} + O\left[ \frac{m_0^2}{r^2}\right] \;,
\end{eqnarray}
where $m_0 = G_N M_0$ is some observable Keplerian mass parameter.
Such an expansion assumes the presence of only power-law deviations from Einstein's theory.
In order to be consistent with it, we shall therefore assume in the present subsection that the Compton wavelength $1/\k$
is much larger than the length scales that are being experimentally probed.

The first equation \eqref{PPNmetric} implies the following second-order expansions for $\Phi$
and its radial derivative $F= \Phi^{\prime}$:
\be
\Phi_{\rm PPN} = -\frac{m_0}{r}+ (\beta-\gamma -1) \frac{m_0^2}{r^2}  + O\left[ \frac{m_0^3}{r^3}\right]
\ee
and
\be \label{FPPN}
F_{\rm PPN} =  \frac{m_0}{r^2}-2(\beta-\gamma-1)\frac{m_0^2}{r^3} + O\left[ \frac{m_0^3}{r^4}\right] \;.  
\ee
Similarly one gets
\be
\Lambda_{\rm PPN} =  \gamma\frac{m_0}{r} + O\left[ \frac{m_0^2}{r^2} \right] \;. \label{LamPPN}
\ee
Let us now compare these expansions to the corresponding $\kappa \to 0$ limits of the torsion bigravity variables
$F= F_1+ F_2$ and $\Lambda_1$. According to \eqref{F1} and \eqref{F2outK0}, we have the following result
\begin{eqnarray}
F_1+F_2 &\underset{\k \to 0}=&\frac{m_F}{r^2}+ \frac{18+44\eta+25\eta^2}{9}\frac{m_1^2}{r^3} \nonumber \\
&& - \frac{4\eta(1+\eta)}{15}\frac{m_1^2R_s^2}{r^5} \label{F2outK0bis} \;,
\end{eqnarray}
where 
\be \label{mF}
m_F=m_1\left(1+\frac{4}{3}\eta\right) -\frac{2}{15}\eta\left(3+4\eta\right)\frac{m_1^2}{R_s} 
\ee
In addition, from Eq. \eqref{Lam_1sol} one gets the following $\k \to 0$ solution for $\Lambda_1$,
\begin{eqnarray}
 \Lambda_1&\underset{\k \to 0}=&\frac{m_{\Lambda}}{r}= \frac{m+\eta \, C_1}{r} \underset{\k \to 0} = \frac{m}{r}\left( 1+\frac{2}{3}\eta \right). \nonumber\\
\end{eqnarray}
One should identify the observable Keplerian mass $m_0$ with the mass parameter $m_F$
(which includes self-gravity effects).
Then one can conclude from the last equality and Eq. \eqref{LamPPN} that we can indeed parametrize the
linearized torsion-bigravity metric by an Eddington $\gamma$ parameter equal to
\be \label{gamma}
\gamma=\frac{m_{\Lambda}}{m_F}=\frac{ 1+\frac{2}{3}\eta }{1+\frac{4}{3}\eta} \;,
\ee
where we consistently neglected the $O(\eta m_1/R_s)$ nonlinear, gravitational binding energy correction term. 

The expression \eqref{gamma} for $\gamma$ encapsulates two main facts related to  a theory involving
both a massless graviton and a massive one. We recall that $\eta$ measures the ratio between the
coupling of the massive graviton to that of the massless one, see Eq. \eqref{cRcFcF2}.
When $\eta \to 0$, $\gamma \to 1$, which is the usual Einstein value, while when
$\eta \to \infty$, $\gamma \to \frac12$, which is the value corresponding to pure massive gravity \cite{Boulware:1973my}.

There are stringent limits on the deviation $\gamma-1$ between the PPN parameter $\gamma$ and its
Einstein value, see notably Refs. \cite{Bertotti:2003rm,Fienga2015}. Note that  the Einstein value $\gamma=1$
is obtained for $\eta \to 0$ and that $\gamma = 1-\frac{2\eta}{3}+O(\eta^2) $ as $\eta \to 0$.
Using the limits from Ref. \cite{Bertotti:2003rm} we see that, in the case where $\k^{-1}$ is very large,
the allowed upper limit on $\eta$ is of order
\be \label{cassini}
\eta \lesssim 10^{-5}\,.
\ee

Coming back to the second-order terms in $F$, Eq. \eqref{F2outK0},
we see that there are two types of deviations from Einstein's theory.
First, there is a term parametrizable by the PPN parameter $\beta$ (see \eqref{FPPN}) with
\be
\frac{18+44\eta+25\eta^2}{9}\frac{m_1^2}{r^3} = - 2(\beta-\gamma-1)\frac{m_0^2}{r^3} \;.
\ee
Using the fact that $m_0= m_F= \left(1+\frac{4}{3}\eta\right) m_1$, we get the following value of
$\beta$ in torsion bigravity
\be
\beta = \frac{18 + 40 \eta + 23 \eta^2}{2 (3 + 4 \eta)^2} \;.
\ee
Note that $\beta \to 23/32$ as $\eta \to \infty$, while, in the limit $\eta \to 0$ we have
\be
\beta = 1-\frac{4\eta}{9}+O(\eta^2)\,.
\ee
Therefore the upper limit \eqref{cassini} on $\eta$ suffices to guarantee that $\beta-1 \lesssim 10^{-5}$,
which is more than sufficient to be compatible with the planetary limits on $\beta-1$ \cite{Fienga2015}.

Concerning the remaining second-order contribution $\propto \eta(1+\eta) m_1^2R_s^2/r^5$ in Eq.  \eqref{F2outK0}, we note
that it is smaller than the non-Einsteinian term $ -2(\beta-\gamma)m_0^2/r^3$ by a factor (when $\eta \to 0$)
of order $(R_s/r)^2$, which is much smaller than 1 in all planetary tests. It can therefore be neglected with
respect to the usual PPN terms. One should take it into account only when discussing relativistic-gravity tests
for near-Earth satellites.

Let us finally recall that the results of the present section have been deduced from the assumption that
the second-order perturbation theory of torsion bigravity yields a sufficiently accurate description
of the deviations from GR. In our conclusions, we will discuss what modifications might
exist if higher-order terms in the perturbation expansion introduce new features in the $\k \to 0$ limit.

\section{Conclusions}

We studied the spherically symmetric (and static) sector of torsion bigravity theories, i.e. 
the four-parameter class of Einstein-Cartan-type theories (with dynamical torsion) that contain 
only two physical excitations (around flat spacetime): 
a massless spin-2 excitation and a massive spin-2 one (of mass $\k$). 
We found that this sector of torsion bigravity has  the
same number of degrees of freedom (as counted by the total differential order of the equations,
after discounting algebraic identities)
as their analogs in {\it ghost-free} bimetric gravity theories, defined \`a la DeRham-Gabadadze-Tolley-Hassan-Rosen
(see Eqs. \eqref{sys3}).
 Knowing that, by contrast, 
spherically symmetric solutions in generic (ghost-full) bimetric gravity theories exhibit one more degree of freedom
(corresponding to the Boulware-Deser ghost), this finding suggests that torsion bigravity might preserve its good ($2+5$)
number of degrees of freedom in the full nonlinear regime.

Another remarkable feature of torsion bigravity concerns its behavior in the limit where the mass of the spin-2 excitation
tends to zero ($ \k \to 0$). Contrary to what happens in all bimetric gravity theories (where ordinary perturbation theory
is marred by the presence of  powers of $\k^{-2}$ that increase at each order of perturbation theory, see, e.g., 
Eqs. \eqref{hBD}, \eqref{Vexp}), we found that the perturbation
theory (around flat space) of torsion bigravity involves no  powers of $\k^{-2}$  at the first two orders of perturbation theory.

We numerically constructed a high-compactness ($|g_{00}+1|_{\rm surface}=0.45$)
(asymptotically flat) star model in torsion bigravity and showed that its  physical properties are significantly
different from those of a  general relativistic star having the same observable Kepler-Schwarzschild mass. See, e.g.,
Eqs. \eqref{CsTBG}, \eqref{CsGR} and equations around. We emphasized that, contrary to the Einstein-Cartan
theory (where the torsion does not propagate), the dynamical torsion present in torsion bigravity 
is generated by the stress-energy tensor $T_{\mu \nu}$ of matter  (even in absence of a spin-density
distribution) and can lead (when $\eta=1$) to significant differences between the Levi-Civita connection, and the
torsionfull one. See Fig. \ref{fig3}.

We also briefly discussed (in section \ref{phenom}) possible phenomenologies of torsion bigravity (depending on the considered
range $\k^{-1}$ of the massive excitation, and on the value of the ratio $\eta$ between the coupling
$G_m$ of the massive graviton to that, $G_0$, of the massless one). As we are not assuming in this work
that an analog of the Vainshtein mechanism might be at work in torsion bigravity, we relied on the fact
that the physical effects of torsion (for non spin-polarized bodies) disappear in the $\eta \to 0$ limit to 
give upper limits on $\eta$  making torsion bigravity compatible with existing solar-system tests of GR. 
We leave to future work an analysis of the compatibility of torsion bigravity with other tests of GR,
notably in binary-pulsar data and gravitational-wave data. 

As already mentioned, remarkable cancellations of $1/\k^2$ factors take place at the first two
orders of the perturbation theory of torsion bigravity. If these cancellations continued at all orders,
one could use torsion bigravity to define an infrared modification of gravity and consider its
cosmological applications (as was already attempted in previous work).
On the other hand, if $1/\k^2$ factors  arise at the third order of perturbation theory,
a preliminary analysis suggests that they could generate contributions to the gravitational acceleration
field $F=\Phi^{\prime}$ (with $g_{00} = - e^{2 \Phi}$) of the type
\be
F_3 \sim    \frac{m^3}{\k^2 r^6}+\frac{m^3}{\k^2 R_s  r^5} \,.
\ee
Compared to the first-order result $F_1 \sim \frac{m}{r^2}$ this would mean that
perturbation theory might lose its validity below a Vainshteinlike radius which could either be
\be
R_V^{(1)} \sim \left(\frac{m^2}{\k^2}\right)^{\frac14} \, \; {\rm or} \; \; R_V^{(2)} \sim \left(\frac{m^2}{\k^2 R_s}\right)^{\frac13} \,.
\ee
If we wished to consider a range $1/\k$ of cosmological magnitude, both possibilities would be problematic
for the phenomenological consequences we deduced above from  second-order perturbation theory.
This would then raise the issue of whether a Vainshteinlike mechanism might be at work in torsion bigravity.
We leave to future work a discussion of this issue, which is expected to be quite different from the 
discussion of the $\k^2 \to 0$ limit in ordinary Fierz-Pauli-type massive gravity models because $\k^2$
enters the torsion bigravity action directly as a denominator (via $c_{F^2} = \frac{\eta \, \lam}{\k^2}$),
while Fierz-Pauli-type actions contain $\k^2$ in the numerator.

We wish, however, to recall that the issue of 
an eventual bad behavior in the $\k^2 \to 0$ limit is separate from the issue of absence
of a sixth degree of freedom, and of ghost-freeness, in the nonlinear regime. In addition, it is only
relevant if one wishes to consider a range $1/\k$ of cosmological magnitude. We are currently
more interested in considering ranges of relevance for modifying the gravitational interaction
of compact objects (neutron stars or black holes).

Our hope is that torsion bigravity might define a theoretically healthy alternative
to GR that could lead to an interesting modified phenomenology for
the gravitational-wave physics of coalescing binary systems of
black holes or neutron stars. The present work is just a first step in this programme.
In particular, we have shown the existence of high-compacness star models.
In the present work, we have only exhibited one model based on an 
unrealistic constant-density equation of state, but we have also constructed 
neutron-star models based on more realistic nuclear equations of state
(with a range $\k^{-1} \sim 10 \; {\rm km}$).

We have also noted that the exterior Schwarzschild solution defines a black hole
solution in torsion bigravity. We leave to future work the issue of whether
this is the unique spherically-symmetric black hole solution of torsion bigravity,
or whether there exist black holes with torsion hair. Our hope is that the different
Young tableau description of the massive gravity field might allow for black-hole hair.

We leave also to future work an Hamiltonian analysis of torsion bigravity to examine
whether its good linearization properties around simple backgrounds, together
with the good degree-of-freedom count in fully nonlinear static-spherically-symmetric solutions,
are sufficient to ensure  ghost-freeness (and mathematical well-posedness) in the full nonlinear theory.

\appendix

\section{Reminders on the  Einstein-Cartan formalism} \label{appA}

In this Appendix we recall some of the basic technicalities of the  Einstein-Cartan(-Weyl-Sciama-Kibble) formalism
(also called Poincar\'e gauge theory).
We generally follow the notation of \cite{Hayashi:1979wj,Hayashi:1980av,Hayashi:1980ir,Hayashi:1980qp},
and of the papers \cite{Nair:2008yh,Nikiforova:2009qr,Nikiforova:2016ngy,Nikiforova:2017saf,Nikiforova:2017xww,Nikiforova:2018pdk}, except for the notation used for the parameters entering into the action.
We use a mostly plus signature and distinguish Lorentz-frame indices ($i, j, k, \ldots= 0,1,2,3 $) from coordinate
ones $\mu, \nu, \ldots=0,1,2,3$. The co-frame (inverse of the vierbein) is denoted $ {e^i}_\mu$ (i.e. 
$g_{\mu \nu} \equiv \eta_{ij} {e^i}_\mu {e^j}_\nu$), while the independent (but metric-preserving) SO$(3,1)$ 
connection is denoted ${A^i}_{ j \mu}$. These fields respectively define the one-forms
$e^i=e^i_{\mu} dx^{\mu}$ and ${{\cA}^i}_{ j}={A^i}_{ j \mu} dx^\mu$. In turn, the basic Cartan formulas
defining the (torsionless) Levi-Civita connection ${\omega^i}_j \equiv {\omega^i}_{j\mu}dx^\mu$ (often
called the Riemannian spin-connection), the Riemann curvature of $e^i$, the torsion  two-form, and curvature two-form
of  ${{\cA}^i}_{ j}$ respectively read:
\be
d e^i + {\omega^i}_j \wedge e^j=0 \; {\rm (vanishing  \, Riemannian \, torsion)}
\ee
\be
{\cR^i}_{ j}=d {\omega^i}_{ j} + {\omega^i}_{ s}\wedge {\omega^s}_{ j} =  \frac12 {R^i}_{ j \mu \nu}dx^\mu \wedge dx^\nu  ,
\ee 
\be
 d e^i + {\cA^i}_j e^j =  -\frac12 {T^i}_{[jk]} e^j \wedge e^k  \,, \label{A-Conn}
\ee
\be
{\cF^i}_{ j}=d {\cA^i}_{ j} + {\cA^i}_{ s}\wedge {\cA^s}_{ j}  = \frac12 {F^i}_{ j \mu \nu}dx^\mu \wedge dx^\nu.
\ee 
The frame components ${T^i}_{[jk]}= - {T^i}_{[kj]}$  of the torsion tensor can be written as 
\be 
T_{i [jk]} = A_{ijk}- A_{ikj} - C_{i [jk]}\,,
\ee
where $C_ {i [jk]}= - C_ {i [kj]}$  are the structure constants of the vierbein, defined as
\be
C_ {i [jk]} \equiv (\d_\mu e_{i\nu} - \d_\nu e_{i\mu})  {e_j}^\mu {e_k}^\nu\; .
\ee
Here, frame indices $i, j, k$ are moved by $\eta_{ij}$.

The explicit links between the contorsion tensor $ K_{ijk} = -  K_{jik}$ (defined as 
$ K_{ijk} \equiv A_{ijk}- \omega_{ijk} $) and the torsion tensor are
\be
K_{ijk}= \frac12 (  T_{i [jk]}+ T_{j [ki]} - T_{k[ij]}) \,,
\ee
\be
T_{i [jk]} =  K_{ijk}- K_{ikj}\,.
\ee

Let us also mention the expression of the Riemannian spin-connection in terms of the vierbein and its derivatives
\be \label{omega=C}
 \omega_{ij\mu} = \omega_{ijk}e^k_\mu =\frac{1}{2}( C_{i [jk]}+ C_{j [ki]} - C_{k[ij]} ) e^k_\mu \; .  
\ee

The frame components of the two curvature tensors,
namely ${R^i}_{jkl} \equiv  {R^i}_{ j \mu \nu} {e_k}^\mu {e_l}^\nu$
and ${F^i}_{jkl} \equiv  {F^i}_{ j \mu \nu} {e_k}^\mu {e_l}^\nu$,
can then be explicitly written 
(in their ``all indices down" forms: $R_{ijkl} \equiv \eta_{i i'}{R^{i'}}_{jkl}$
and $F_{ijkl} \equiv \eta_{i i'}{F^{i'}}_{jkl}$) as
\bea
R_{ijkl} &=& {e_k}^\mu {e_l}^\nu \l \partial_\mu \omega_{ij\nu} - \partial_\nu \omega_{ij\mu} \right.\\ \nonumber
&+& \left. \eta^{mn} \omega_{im\mu}\omega_{n j\nu} - \eta^{mn} \omega_{im\nu}\omega_{n j\mu} \r \; ,
\eea
\bea
F_{ijkl} &=& {e_k}^\mu {e_l}^\nu \l \partial_\mu A_{ij\nu} - \partial_\nu A_{ij\mu} \right.\\ \nonumber
&+& \left. \eta^{mn} A_{im\mu}A_{n j\nu} - \eta^{mn} A_{im\nu}A_{n j\mu} \r \; .
\eea
The tensor and scalar curvatures with contracted indices are defined as follows, 
\be 
R_{ij}=\eta^{kl}R_{kilj} =\eta^{kl}R_{ikjl} \;,\;\; R=\eta^{ij}R_{ij} \; , 
\ee
\be 
F_{ij}=\eta^{kl}F_{kilj} =\eta^{kl}F_{ikjl} \;,\;\; F=\eta^{ij}F_{ij} \; .
\ee

\section{Link with the notation used in \cite{Nikiforova:2018pdk}}\label{appB}

In our previous paper Ref. \cite{Nikiforova:2018pdk} we considered the more general action
\bea  \label{lagrangianold}
L &=& \frac{3}{2}  \widetilde{\alpha}\, F[e, A, \d A]
+ \frac{3}{2} \overline\alpha \, R[e, \d e, \d^2 e]  + c_2 \\ \nonumber
&+& c_3 F^{ij}F_{ij}
 + c_4 F^{ij}F_{ji} + c_5 F^2 + c_6 (\epsilon^{ijkl}  F_{ijkl})^2 \; ,  
 \eea
containing a cosmological constant $c_2$ and five coupling constants 
$ \widetilde{\alpha}, \overline\alpha,  c_3, c_4, c_5, c_6$. 
In order to avoid pathologies around flat spacetime, these parameters must satisfy
the equation
\be
c_3+c_4=-3 \, c_5 \; , \label{c3c4c5}
\ee 
and the inequalities
\be \label{ineqs}
\tal > 0, \;\;\;\; \bal > 0, \;\;\;\; c_5 < 0, \;\;\;\; c_6 > 0 \,.
\ee

The field content of such a model around  flat space consists of a massless spin-2, a massive spin-2 and a massive pseudoscalar field. The corresponding masses are \cite{Nikiforova:2009qr}
\be \label{m2}
m_2^2= \k^2=\frac{\tal (\tal  + \bal)}{2  \bal \, (- c_5)}  >0 \,,
\ee
 while that of the pseudoscalar field is
\be \label{m0}
m^2_0=\frac{\widetilde{\alpha}}{16c_6} >0\;.  
\ee
We define torsion bigravity by setting $c_6=0$ so as to ``freeze out" the pseudoscalar field (which becomes
infinitely massive). We also set for simplicity the bare cosmological constant $c_2$ to zero.
This leaves us with only four independent parameters: $\tal$, $\bal$, $c_3$ and $c_4$.

We then find convenient to change the notation of the parameters and to introduce
\bea
c_F &\equiv& \frac32 \tal  \, ; \, c_R \equiv  \frac32 \bal  \, ; \nonumber\\
 c_{F^2} &\equiv&  c_3+c_4 = - 3 \, c_5 \, ; \, c_{34} \equiv c_3 - c_4\,.
\eea
In terms of these parameters, and of the symmetric ($F_{(ij)}$) and antisymmetric ($F_{[ij]}$) parts
of $F_{ij}=F_{(ij)}+F_{[ij]}$, the torsion bigravity Lagrangian density reads
\bea  \label{lagrangian2}
L_{\rm TBG} &=& c_R \, R[e, \d e, \d^2 e] +c_F \, F[e, A, \d A] \\ \nonumber
&+& c_{F^2}\left( F_{(ij)} F^{(ij)} - \frac13 F^2 \right)
 + c_{34} F_{[ij]} F^{[ij]}  \; . 
 \eea
 This model contains only a massless spin-2 and a massive one of squared mass
 \be
 \k^2 = \frac{\eta \, \lam}{c_{F^2}}\,.
 \ee
 
 \section*{Acknowledgments} 
The authors thank Cedric Deffayet, J\"urg Fr\"ohlich, Patrick Iglesias-Zemmour, Sergiu Klainerman and Mikhail Volkov for useful discussions.



\begin{thebibliography}{20}

\bibitem{Touboul:2017grn} 
  P.~Touboul {\it et al.},
  ``MICROSCOPE Mission: First Results of a Space Test of the Equivalence Principle,''
  Phys.\ Rev.\ Lett.\  {\bf 119}, no. 23, 231101 (2017)
  [arXiv:1712.01176 [astro-ph.IM]].
  
\bibitem{LIGOScientific:2019fpa} 
  B.~P.~Abbott {\it et al.} [LIGO Scientific and Virgo Collaborations],
  ``Tests of General Relativity with the Binary Black Hole Signals from the LIGO-Virgo Catalog GWTC-1,''
  arXiv:1903.04467 [gr-qc].
  
\bibitem{Abbott:2018lct} 
  B.~P.~Abbott {\it et al.} [LIGO Scientific and Virgo Collaborations],
``Tests of General Relativity with GW170817,''
  arXiv:1811.00364 [gr-qc].
  
\bibitem{Tanabashi:2018oca} 
  M.~Tanabashi {\it et al.} [Particle Data Group],
  Phys.\ Rev.\ D {\bf 98}, no. 3, 030001 (2018).
  doi:10.1103/PhysRevD.98.030001
  
\bibitem{Capozziello:2011et} 
  S.~Capozziello and M.~De Laurentis,
  ``Extended Theories of Gravity,''
  Phys.\ Rept.\  {\bf 509}, 167 (2011)
  [arXiv:1108.6266 [gr-qc]].
  
\bibitem{Berti:2015itd} 
  E.~Berti {\it et al.},
  ``Testing General Relativity with Present and Future Astrophysical Observations,''
  Class.\ Quant.\ Grav.\  {\bf 32}, 243001 (2015)
  [arXiv:1501.07274 [gr-qc]].
  
\bibitem{Cartan:1923zea} 
  E.~Cartan,
  ``Sur les vari\'et\'es \`a connexion affine et la th\'eorie de la relativit\'e g\'en\'eralis\'ee. (premi\`ere partie),''
  Annales Sci.\ Ecole Norm.\ Sup.\  {\bf 40}, 325 (1923).
  
\bibitem{Cartan:1924yea} 
  E.~Cartan,
  ``Sur les vari\'et\'es \`a connexion affine et la th\'eorie de la relativit\'e g\'en\'eralis\'ee. (Suite).,''
  Annales Sci.\ Ecole Norm.\ Sup.\  {\bf 41}, 1 (1924).
  
\bibitem{Cartan1925}   
E.~Cartan,
  ``Sur les vari\'et\'es \`a connexion affine et la th\'eorie de la relativit\'e g\'en\'eralis\'ee. 
  (deuxi\`eme partie)),''
  Annales Sci.\ Ecole Norm.\ Sup.\  {\bf 42}, 17 (1925).
  
 \bibitem{Sciama1962}
 D.~W., Sciama, 
 ``On the analogy between charge and spin in general relativity",
 in {\it Recent Developments in General Relativity}, (Pergamon Press, Oxford, 1962), pp. 415--439. 
 
  
\bibitem{Kibble:1961ba} 
  T.~W.~B.~Kibble,
  ``Lorentz invariance and the gravitational field,''
  J.\ Math.\ Phys.\  {\bf 2}, 212 (1961).
  
\bibitem{Hehl:1976kj} 
  F.~W.~Hehl, P.~Von Der Heyde, G.~D.~Kerlick and J.~M.~Nester,
  ``General Relativity with Spin and Torsion: Foundations and Prospects,''
  Rev.\ Mod.\ Phys.\  {\bf 48}, 393 (1976).
  
\bibitem{Freedman:1976xh} 
  D.~Z.~Freedman, P.~van Nieuwenhuizen and S.~Ferrara,
  ``Progress Toward a Theory of Supergravity,''
  Phys.\ Rev.\ D {\bf 13}, 3214 (1976).
  
\bibitem{Deser:1976eh} 
  S.~Deser and B.~Zumino,
  ``Consistent Supergravity,''
  Phys.\ Lett.\ B {\bf 62}, 335 (1976)
  [Phys.\ Lett.\  {\bf 62B}, 335 (1976)].
  
\bibitem{Stelle:1976gc} 
  K.~S.~Stelle,
  ``Renormalization of Higher Derivative Quantum Gravity,''
  Phys.\ Rev.\ D {\bf 16}, 953 (1977).
  
\bibitem{Stelle:1977ry} 
  K.~S.~Stelle,
  ``Classical Gravity with Higher Derivatives,''
  Gen.\ Rel.\ Grav.\  {\bf 9}, 353 (1978).
  
\bibitem{Neville:1978bk} 
  D.~E.~Neville,
  ``A Gravity Lagrangian With Ghost Free Curvature**2 Terms,''
  Phys.\ Rev.\ D {\bf 18}, 3535 (1978).
 
\bibitem{Neville:1979rb} 
  D.~E.~Neville,
  ``Gravity Theories With Propagating Torsion,''
  Phys.\ Rev.\ D {\bf 21}, 867 (1980).
  
\bibitem{Sezgin:1979zf} 
  E.~Sezgin and P.~van Nieuwenhuizen,
  ``New Ghost Free Gravity Lagrangians with Propagating Torsion,''
  Phys.\ Rev.\ D {\bf 21}, 3269 (1980).
  
\bibitem{Sezgin:1981xs} 
  E.~Sezgin,
  ``Class of Ghost Free Gravity Lagrangians With Massive or Massless Propagating Torsion,''
  Phys.\ Rev.\ D {\bf 24}, 1677 (1981).
  
\bibitem{Hayashi:1979wj} 
  K.~Hayashi and T.~Shirafuji,
  ``Gravity from Poincare Gauge Theory of the Fundamental Particles. 1. General Formulation,''
  Prog.\ Theor.\ Phys.\  {\bf 64}, 866 (1980)
  Erratum: [Prog.\ Theor.\ Phys.\  {\bf 65}, 2079 (1981)].
 
\bibitem{Hayashi:1980av} 
  K.~Hayashi and T.~Shirafuji,
  ``Gravity From Poincare Gauge Theory Of The Fundamental Particles. 2. Equations Of Motion For Test Bodies And Various Limits,''
  Prog.\ Theor.\ Phys.\  {\bf 64}, 883 (1980)
  Erratum: [Prog.\ Theor.\ Phys.\  {\bf 65}, 2079 (1981)].
  
\bibitem{Hayashi:1980ir} 
  K.~Hayashi and T.~Shirafuji,
  ``Gravity From Poincare Gauge Theory of the Fundamental Particles. 3. Weak Field Approximation,''
  Prog.\ Theor.\ Phys.\  {\bf 64}, 1435 (1980)
  Erratum: [Prog.\ Theor.\ Phys.\  {\bf 66}, 741 (1981)].
  
\bibitem{Hayashi:1980qp} 
  K.~Hayashi and T.~Shirafuji,
  ``Gravity From Poincare Gauge Theory of the Fundamental Particles. 4. Mass and Energy of Particle Spectrum,''
  Prog.\ Theor.\ Phys.\  {\bf 64}, 2222 (1980).
  
\bibitem{Nair:2008yh} 
  V.~P.~Nair, S.~Randjbar-Daemi and V.~Rubakov,
  ``Massive Spin-2 fields of Geometric Origin in Curved Spacetimes,''
  Phys.\ Rev.\ D {\bf 80}, 104031 (2009)
  [arXiv:0811.3781 [hep-th]].
  
\bibitem{Nikiforova:2009qr} 
  V.~Nikiforova, S.~Randjbar-Daemi and V.~Rubakov,
  ``Infrared Modified Gravity with Dynamical Torsion,''
  Phys.\ Rev.\ D {\bf 80}, 124050 (2009)
  [arXiv:0905.3732 [hep-th]].
  
   
\bibitem{Fierz:1939zz} 
  M.~Fierz,
  ``Force-free particles with any spin,''
  Helv.\ Phys.\ Acta {\bf 12}, 3 (1939).
  
\bibitem{Pauli:1939xp} 
  W.~Pauli and M.~Fierz,
  ``On Relativistic Field Equations of Particles With Arbitrary Spin in an Electromagnetic Field,''
  Helv.\ Phys.\ Acta {\bf 12}, 297 (1939).
  
\bibitem{Fierz:1939ix} 
  M.~Fierz and W.~Pauli,
  ``On relativistic wave equations for particles of arbitrary spin in an electromagnetic field,''
  Proc.\ Roy.\ Soc.\ Lond.\ A {\bf 173}, 211 (1939).

  
\bibitem{vanDam:1970vg} 
  H.~van Dam and M.~J.~G.~Veltman,
  ``Massive and massless Yang-Mills and gravitational fields,''
  Nucl.\ Phys.\ B {\bf 22}, 397 (1970).
  
\bibitem{Zakharov:1970cc} 
  V.~I.~Zakharov,
  ``Linearized gravitation theory and the graviton mass,''
  JETP Lett.\  {\bf 12}, 312 (1970)
  [Pisma Zh.\ Eksp.\ Teor.\ Fiz.\  {\bf 12}, 447 (1970)].
  
\bibitem{Vainshtein:1972sx} 
  A.~I.~Vainshtein,
  ``To the problem of nonvanishing gravitation mass,''
  Phys.\ Lett.\  {\bf 39B}, 393 (1972).
  
\bibitem{Boulware:1973my} 
  D.~G.~Boulware and S.~Deser,
  ``Can gravitation have a finite range?,''
  Phys.\ Rev.\ D {\bf 6}, 3368 (1972).

\bibitem{deRham:2010kj} 
  C.~de Rham, G.~Gabadadze and A.~J.~Tolley,
  ``Resummation of Massive Gravity,''
  Phys.\ Rev.\ Lett.\  {\bf 106}, 231101 (2011)
  doi:10.1103/PhysRevLett.106.231101
  [arXiv:1011.1232 [hep-th]].
  
\bibitem{Hassan:2011zd} 
  S.~F.~Hassan and R.~A.~Rosen,
  ``Bimetric Gravity from Ghost-free Massive Gravity,''
  JHEP {\bf 1202}, 126 (2012)
  [arXiv:1109.3515 [hep-th]].
  
\bibitem{Curtright:1980yk} 
  T.~Curtright,
  ``Generalized Gauge Fields,''
  Phys.\ Lett.\  {\bf 165B}, 304 (1985).
  
\bibitem{Curtright:1980yj} 
  T.~L.~Curtright and P.~G.~O.~Freund,
  ``Massive Dual Fields,''
  Nucl.\ Phys.\ B {\bf 172}, 413 (1980).
  
\bibitem{Nikiforova:2016ngy} 
  V.~Nikiforova, S.~Randjbar-Daemi and V.~Rubakov,
  ``Self-accelerating Universe in modified gravity with dynamical torsion,''
  Phys.\ Rev.\ D {\bf 95}, no. 2, 024013 (2017)
  [arXiv:1606.02565 [hep-th]].
  
\bibitem{Nikiforova:2017saf} 
  V.~Nikiforova,
  ``The stability of self-accelerating Universe in modified gravity with dynamical torsion,''
  Int.\ J.\ Mod.\ Phys.\ A {\bf 32}, no. 23n24, 1750137 (2017)
  [arXiv:1705.00856 [hep-th]].
  
\bibitem{Nikiforova:2017xww} 
  V.~Nikiforova,
 ``Stability of self-accelerating Universe in modified gravity with dynamical torsion: the case of small background torsion,''
  Int.\ J.\ Mod.\ Phys.\ A {\bf 33}, no. 07, 1850039 (2018)
  [arXiv:1711.03718 [hep-th]].
   
  
\bibitem{Nikiforova:2018pdk} 
  V.~Nikiforova and T.~Damour,
  ``Infrared modified gravity with propagating torsion: instability of torsionfull de Sitter-like solutions,''
  Phys.\ Rev.\ D {\bf 97}, no. 12, 124014 (2018)
  [arXiv:1804.09215 [gr-qc]].
  
\bibitem{Zhang:1982jn} 
  Y.~Z.~Zhang,
  ``Approximate Solutions for General Riemann-Cartan Type $R+R^2$ Theories of Gravitation,''
  Phys.\ Rev.\ D {\bf 28}, 1866 (1983).

  
\bibitem{Afonso:2017bxr} 
  V.~I.~Afonso, C.~Bejarano, J.~Beltran Jimenez, G.~J.~Olmo and E.~Orazi,
  ``The trivial role of torsion in projective invariant theories of gravity with non-minimally coupled matter fields,''
  Class.\ Quant.\ Grav.\  {\bf 34}, no. 23, 235003 (2017)
  [arXiv:1705.03806 [gr-qc]].
  
\bibitem{BeltranJimenez:2017doy} 
  J.~Beltran Jimenez, L.~Heisenberg, G.~J.~Olmo and D.~Rubiera-Garcia,
  ``BornÐInfeld inspired modifications of gravity,''
  Phys.\ Rept.\  {\bf 727}, 1 (2018)
  [arXiv:1704.03351 [gr-qc]].
  
\bibitem{Rauch:1981tva} 
  R.~Rauch and H.~T.~Nieh,
  ``Birkhoff's Theorem for General {Riemann-Cartan} Type $R+R^2$ Theories of Gravity,''
  Phys.\ Rev.\ D {\bf 24}, 2029 (1981).
  
   
\bibitem{Babichev:2009us} 
  E.~Babichev, C.~Deffayet and R.~Ziour,
  ``The Vainshtein mechanism in the Decoupling Limit of massive gravity,''
  JHEP {\bf 0905}, 098 (2009)
  [arXiv:0901.0393 [hep-th]].
  
\bibitem{Damour:2002gp} 
  T.~Damour, I.~I.~Kogan and A.~Papazoglou,
  ``Spherically symmetric space-times in massive gravity,''
  Phys.\ Rev.\ D {\bf 67}, 064009 (2003)
  [hep-th/0212155].
  
\bibitem{Deffayet:2005ys} 
  C.~Deffayet and J.~W.~Rombouts,
  ``Ghosts, strong coupling and accidental symmetries in massive gravity,''
  Phys.\ Rev.\ D {\bf 72}, 044003 (2005)
  [gr-qc/0505134].
  
\bibitem{Creminelli:2005qk} 
  P.~Creminelli, A.~Nicolis, M.~Papucci and E.~Trincherini,
  ``Ghosts in massive gravity,''
  JHEP {\bf 0509}, 003 (2005)
  [hep-th/0505147].
  
\bibitem{Babichev:2010jd} 
  E.~Babichev, C.~Deffayet and R.~Ziour,
  ``The Recovery of General Relativity in massive gravity via the Vainshtein mechanism,''
  Phys.\ Rev.\ D {\bf 82}, 104008 (2010)
  [arXiv:1007.4506 [gr-qc]].
  
\bibitem{Volkov:2012wp} 
  M.~S.~Volkov,
  ``Hairy black holes in the ghost-free bigravity theory,''
  Phys.\ Rev.\ D {\bf 85}, 124043 (2012)
  [arXiv:1202.6682 [hep-th]].
  
   
\bibitem{Deffayet:2001uk} 
  C.~Deffayet, G.~R.~Dvali, G.~Gabadadze and A.~I.~Vainshtein,
  ``Nonperturbative continuity in graviton mass versus perturbative discontinuity,''
  Phys.\ Rev.\ D {\bf 65}, 044026 (2002)
  [hep-th/0106001].
  
  
\bibitem{Fischbach:1999bc} 
  E.~Fischbach and C.~L.~Talmadge,
  ``The search for nonNewtonian gravity,''
  New York, USA: Springer (1999) 305 p
  
\bibitem{Adelberger:2003zx} 
  E.~G.~Adelberger, B.~R.~Heckel and A.~E.~Nelson,
  ``Tests of the gravitational inverse square law,''
  Ann.\ Rev.\ Nucl.\ Part.\ Sci.\  {\bf 53}, 77 (2003)
  [hep-ph/0307284].
  
\bibitem{Weinberg:1972kfs} 
  S.~Weinberg,
  ``Gravitation and Cosmology : Principles and Applications of the General Theory of Relativity,''
  (John Wiley and sons, New York, 1972).

  
\bibitem{Bertotti:2003rm} 
  B.~Bertotti, L.~Iess and P.~Tortora,
  ``A test of general relativity using radio links with the Cassini spacecraft,''
  Nature {\bf 425}, 374 (2003).
 
 \bibitem{Fienga2015} 
 A. Fienga et al., 
 ``Numerical estimation of the sensitivity of INPOP planetary ephemerides to general relativity parameters,"
 Cel. Mech. Dyn. Astr. {\bf 123}, 325 (2015). 

\bibitem{Schwarzschild:1916ae} 
  K.~Schwarzschild,
  ``On the gravitational field of a sphere of incompressible fluid according to Einstein's theory,''
  Sitzungsber.\ Preuss.\ Akad.\ Wiss.\ Berlin (Math.\ Phys.\ ) {\bf 1916}, 424 (1916)
  [physics/9912033 [physics.hist-ph]].
  
\bibitem{Schwarzschild:1916uq} 
  K.~Schwarzschild,
  ``On the gravitational field of a mass point according to Einstein's theory,''
  Sitzungsber.\ Preuss.\ Akad.\ Wiss.\ Berlin (Math.\ Phys.\ ) {\bf 1916}, 189 (1916)
  [physics/9905030].
  
  \bibitem{Damour:2009sm} 
  T.~Damour,
  ``Gravitational Self Force in a Schwarzschild Background and the Effective One Body Formalism,''
  Phys.\ Rev.\ D {\bf 81}, 024017 (2010)
  [arXiv:0910.5533 [gr-qc]].

  
\end{thebibliography}
\end{document}